\documentclass[10pt,journal,final,twocolumn]{IEEEtran}

\IEEEoverridecommandlockouts

\usepackage{graphicx,array}
\usepackage{amssymb,amsmath,enumerate,empheq,fancybox}
\usepackage{cite}
\usepackage{algorithm}
\usepackage{algpseudocode}

\usepackage{tikz}
\usetikzlibrary{calc,shapes,arrows,petri}

\usepackage{caption,subcaption}

\newtheorem{theorem}{\bf Theorem}
\newtheorem{assum}{\bf Assumption}
\newtheorem{prop}{\bf Proposition}
\newtheorem{defn}{\bf Definition}
\newtheorem{lemma}{\bf Lemma}
\newtheorem{corollary}{\bf Corollary}
\newtheorem{remark}{\bf Remark}

\def\be{\begin{equation}}
\def\ee{\end{equation}}
\def\ben{\begin{equation*}}
\def\een{\end{equation*}}
\newcommand{\dfb}{\stackrel{\Delta}{=}}

\def\r{\mathbb{R}}

\def\vp{\varphi}

\def\ae{\varkappa}

\def\u{\mathfrak U}

\def\vk{\varkappa}

\title{Lyapunov Event-triggered Stabilization\\ with a Known Convergence Rate}

\author{Anton V. Proskurnikov,~\IEEEmembership{Senior Member,~IEEE} and  Manuel Mazo Jr.,~\IEEEmembership{Senior Member,~IEEE}
\thanks{The authors are with Delft Center for Systems and Control, Delft University of Technology, The Netherlands.
{E-mail: \tt\small anton.p.1982@ieee.org; m.mazo@tudelft.nl}}%
\thanks{The work is supported by NWO Domain TTW, Netherlands, under the project TTW\#13712 ``From Individual Automated Vehicles to Cooperative Traffic Management -- predicting the benefits of automated driving through on-road human behavior assessment and traffic flow models'' (IAVTRM). 
It is also partially funded by the project SENTIENT funded by the European Research Council (ERC) under the EU's Horizon 2020 research and 
innovation programme (ERC-StG~2017, \#755953).
}
\thanks{A special case of Theorem~\ref{thm.dwell} (dealing with exponentially stabilizing CLF)
was presented~\cite{ProMazo_HSCC} on the $21^{\rm st}$ ACM Conference on Hybrid Systems Computation and Control (HSCC'2018), Porto, Portugal, April 11-13, 2018.}
}

\begin{document}

\maketitle

\begin{abstract}

A constructive tool of nonlinear control systems design, the method of Control Lyapunov Functions (CLF) has found numerous applications in stabilization problems for continuous-time, discrete-time and hybrid systems.
In this paper, we address the fundamental question: given a CLF, corresponding to the continuous-time controller with some predefined (e.g. exponential) convergence rate, can the same convergence rate be provided by an event-triggered controller? Under certain assumptions, we give an affirmative answer to this question and show that the corresponding event-based controllers provide positive dwell-times between the consecutive events. Furthermore, we prove the existence of self-triggered and periodic event-triggered controllers, providing stabilization with a known convergence rate.
\end{abstract}


\section{Introduction}

The seminal idea to use the second Lyapunov method as a tool of control \emph{design}~\cite{KalmanBertram:1960} has naturally lead to the idea of control Lyapunov Function (CLF). A CLF is a function that becomes a Lyapunov function of the closed-loop system under an appropriate (usually, non-unique) choice of the controller.
The fundamental Artstein theorem~\cite{Artstein:1983} states that the existence of a CLF is necessary and sufficient for stabilization
of a general nonlinear system by a ``relaxed'' controller, mapping the system's state into a probability measure.
For an \emph{affine} unconstrained system, a usual static stabilizing controller can always be found, as shown in the seminal work~\cite{Sontag:1989}.

In general, to find a CLF for a given control system is a non-trivial problem since the set of CLFs may have a very sophisticated structure,
e.g. be disconnected~\cite{Rantzer:2001}. However, in some important situations a CLF can be explicitly found. Examples include
some homogeneous systems~\cite{FauborgPomet:2000}, feedback-linearizable, passive or feedback-passive systems~\cite{KokotovicArcak:2001,Khalil} and cascaded systems~\cite{Praly:1991}, for which both CLFs and stabilizing controllers can be
delivered by the \emph{backstepping} and \emph{forwarding} procedures~\cite{KrsticKokotovicBook,SepulchreKokotovicBook}. The CLF method has recently been empowered by the development of algorithms and software for convex optimization~\cite{WangLiu:2017,Furqon:2017} and genetic programming~\cite{VerdierMazo:2017}.

Nowadays the method of CLF is recognized as a powerful tool in nonlinear control systems design~\cite{KrsticKokotovicBook,Khalil,SepulchreKokotovicBook}.
A CLF gives a solution to the Hamilton-Jacobi-Bellman equation for an appropriate performance index, giving a solution to the \emph{inverse optimality} problem~\cite{FreemanKokotovic:1996}. Another numerical method to compute CLFs~\cite{Camilli:2008} employs the so-called Zubov equation. The method of CLF has been extended to uncertain~\cite{FreemanKokotovic:1996,FreemanKokotovicBook}, discrete-time~\cite{KellettTeel:2004}, time-delay~\cite{Jancovic:2001} and hybrid systems~\cite{Sanfelice:2013,Ames:2014}. Combining CLFs and Control Barrier Functions (CBFs), correct-by-design controllers for stabilization of \emph{constrained} (``safety-critical'') systems have been proposed~\cite{RomdlonyJayawardhana:2016,NilssonTabuada:2016,AmesTabuada:2017}.

For continuous-time systems, CLF-based controllers are also continuous-time. Their implementation on digital platforms requires to introduce time sampling. The simplest approach is based on \emph{emulation} of the continuous-time feedback by a discrete-time control, sampled at a high rate. Rigorous stability analysis of the resulting sampled-time systems is highly non-trivial; we refer the reader to~\cite{Hetel2017} for a detailed survey of the existing methods. A more general framework to sample-time control design, based on a direct \emph{discretization} of the nonlinear control system and approximating it by a nonlinear discrete-time inclusion, has been developed in~\cite{NesicTeelKokotovich:1999,NesicTeel:2004,ArcakNesic:2004}. This method allows to
design controllers that cannot be directly redesigned from continuous-time algorithms, but the relevant design procedures and stability analysis are sophisticated.

The necessity to use communication, computational and power resources parsimoniously has motivated to study digital controllers that are based on \emph{event-triggered} sampling, which
has a number of advantages over classical time-triggered control~\cite{Astrom:2002,Tabuada:2007,BorgersHeemels:2014,Araujo:2014,Postoyan2015}.
Event-triggered control strategies can be efficiently analyzed by using the theories of hybrid systems~\cite{Goebel:2009,Postoyan2015,DolkBorgersHeemels:2017}, switching systems~\cite{SelivanovFridman:2016}, delayed systems~\cite{YueTianHan:2013,SelivanovFridman:2016-1} and impulsive systems~\cite{LiuLiuDou:2014}. It should be noticed that the event-triggered sampling is aperiodic and, unlike the classical time-triggered designs, the inter-sampling interval need not necessarily be sufficiently small: the control can be frozen for a long time, provided that the behavior of the system is satisfactory and requires no intervention. On the other hand, with event-triggered sampling one has to prove the existence of positive dwell time between consecutive
events: even though mathematically any non-Zeno trajectory is admissible, in real-time control systems the sampling rate is always limited.

A natural question arises whether the existence of a CLF makes it possible to design an event-triggered controller. In a few situations, the answer is known to be affirmative.
The most studied is the case where the CLF appears to be a so called \emph{ISS Lyapunov} function~\cite{Tabuada:2007,Postoyan2015} and allows to prove the \emph{input-to-state stability} (ISS)
of the closed-loop system with respect to measurement errors. A more recent result from~\cite{SeuretPrieurMarchand:2013} relaxes the ISS condition to a stronger version of usual asymptotic stability, however the control algorithm from~\cite{SeuretPrieurMarchand:2013}, in general, does not ensure the absence of Zeno solutions.
Another approach, based on Sontag's universal formula~\cite{Sontag:1989} has been proposed in~\cite{Marchand:2013,Marchand:2013IFAC}.
All of these results impose limitations, discussed in detail in Section~\ref{sec.prelim}. In particular, the estimation of the convergence rate for the methods proposed in~\cite{SeuretPrieurMarchand:2013,Marchand:2013,Marchand:2013IFAC} is a non-trivial problem.
In many situations a CLF can be designed that provides some known convergence rate (e.g. exponentially stabilizing CLFs~\cite{ProMazo_HSCC,Ames:2014}) in continuous time.
A natural question arises whether \emph{event-based} controllers can provide the same (or an arbitrarily close) convergence rate.
In this paper, we give an affirmative answer to this fundamental question. Under natural assumptions, we design an event-triggered controller, providing a known convergence rate and a positive \emph{dwell time} between consecutive events.
Furthermore, we design self-triggered and periodic event-triggered controllers that simplify real-time task scheduling.

The paper is organized as follows. Section~\ref{sec.prelim} gives the definition of CLF and related concepts and sets up the problem of event-triggered stabilization with a predefined convergence rate. The solution to this problem, being the main result of the paper, is offered in Section~\ref{sec.event}, where event-triggered, self-triggered and periodic event-triggered stabilizing controllers are designed. In Section~\ref{sec.example}, the main results are illustrated by numerical examples. Section~\ref{sec.concl} concludes the paper. Appendix contains some technical proofs and discussion on the key assumption in the main result. 
\section{Preliminaries and problem setup}\label{sec.prelim}

Henceforth $\r^{m\times n}$ stands for the set of $m\times n$ real matrices, $\r^n=\r^{n\times 1}$.
Given a function $G:\r^n\to\r^m$ that maps $x\in\r^n$ into $G(x)=(G_1(x),\ldots,G_m(x))^{\top}\in\r^m$,
we use $G'(x)=\big(\frac{\partial G_i(x)}{\partial x_j}\big)\in\r^{m\times n}$ to denote its Jacobian matrix.

\subsection{Control Lyapunov functions in stabilization problems}

To simplify matters, henceforth we deal with the problem of global asymptotic stabilization. Consider the following control system
\be\label{eq.syst}
\dot x(t)=F(x(t),u(t)),\quad t\ge 0,
\ee
where $x(t)\in\r^{d}$ stands for the state vector and $u(t)\in U\subseteq\r^{m}$ is the control input (the case $U=\r^m$ corresponds to the absence of input constraints).
Our goal is to find a controller $u(\cdot)=\mathcal U(x(\cdot))$, where $\mathcal U:x(\cdot)\mapsto u(\cdot)$ is some \emph{causal} (non-anticipating) operator, such that for any $x(0)\in\r^{d}$ the solution to the closed-loop system is forward complete
(exists up to $t=+\infty$) and converges to the unique equilibrium $x=0$
\be\label{eq.stab}
x(t)\xrightarrow[t\to\infty]{} 0\quad\forall x(0)\in\r^d,\quad F(0,\mathcal U(0))=0.
\ee

We now give the definition of CLF. Following~\cite{Sontag:1989}, we henceforth assume CLFs to be smooth, radially unbounded (or \emph{proper}) and positive definite.
\begin{defn}\cite{Sontag:1989}
A $C^1$-smooth function $V:\r^d\to\r$ is called a \emph{control Lyapunov function} (CLF) 
\begin{gather}
V(0)=0,\quad V(x)>0\,\forall x\ne 0,\quad \lim_{|x|\to\infty}V(x)=\infty;\label{eq.pos-def}\\
\inf_{u\in U} V'(x)F(x,u)<0\quad\forall x\ne 0.\label{eq.inf-u}
\end{gather}
\end{defn}
\vskip 1mm
\par The condition~\eqref{eq.inf-u}, obviously, can be reformulated as follows
\be\label{eq.inf-u-2}
\forall x\ne 0\,\exists u(x)\in U\;\text{such that $V'(x)F(x,u(x))<0$.}
\ee
If $F(x,u)$ is Lebesgue measurable (e.g., continuous), then the set $\{x\ne 0,u\in U:V'(x)F(x,u)<0\}$ is also measurable and the Aumann measurable selector theorem~\cite[Theorem~5.2]{Himmelberg:1975} implies that the function $u(x)$ can be chosen measurable; however, it can be discontinuous and infeasible (the closed-loop system has no solution for some initial condition).
Some systems~\eqref{eq.syst} with continuous right-hand sides cannot be stabilized by usual controllers in spite of the existence of a CLF, however, they can be stabilized by
a \emph{``relaxed''} control~\cite{Artstein:1983} $x\mapsto v(x)$, where $v(x)$ is a probability distribution on $U$.

The situation becomes much simpler in the case of \emph{affine} system~\eqref{eq.syst} with $F(x,u)=f(x)+g(x)u$. Assuming that $f:\r^d\to\r^d$ and $g:\r^d\to\r^{d\times m}$ are continuous and $U$ is convex, the existence of a CLF ensures the possibility to design a controller $u=u(x)$, where $u:\r^d\to U$ is \emph{continuous} everywhere except for, possibly, $x=0$~\cite{Artstein:1983}. While the original proof from~\cite{Artstein:1983} was not fully constructive, Sontag~\cite{Sontag:1989} has proposed an explicit universal formula, giving a broad class of stabilizing controllers. Assuming that $U=\r^m$, let
\ben
a(x)\dfb V'(x)f(x),\quad b(x)\dfb V'(x)g(x).
\een
Then~\eqref{eq.inf-u} means that $a(x)<0$ whenever $b(x)=0$ and $x\ne 0$. In the scalar case ($m=1$), Sontag's controller is
\be\label{eq.sontag-scal}
u(x)=
\begin{cases}
-\frac{a(x)+\sqrt{a(x)^2+q(b(x))b(x)}}{b(x)},\,&b(x)>0\\
0,&\text{otherwise.}
\end{cases}
\ee
Here $q(b)$ is a continuous function, $q(0)=0$. It is shown~\cite{Sontag:1989} that the control~\eqref{eq.sontag-scal} is continuous at any $x\ne 0$, moreover, if $a(\cdot)$, $b(\cdot)$ and $q(\cdot)$ are $C^k$-smooth (respectively, real analytic), the same holds for $u(\cdot)$ in the domain $\r^d\setminus\{0\}$. The global continuity requires an addition ``small control'' property~\cite{Sontag:1989}. Similar controllers
have been found for a more general case, where $m>1$ and $U$ is a closed ball in $\r^m$~\cite{Sontag:1991}.

\subsection{CLF and event-triggered control}

Dealing with continuous-time systems~\eqref{eq.syst}, the CLF-based controller $u=\u(x)$ is also continuous-time, and its implementation on digital platforms requires time-sampling. Formally, the control command is computed and sent to the plant at time instants $t_0=0<t_1<\ldots<t_n<\ldots$ and remain constant $u(t)\equiv u_n$ for $t\in [t_n,t_{n+1})$.
The approach broadly used in engineering is to emulate the continuous-time feedback by sufficiently fast periodic or aperiodic sampling (the intervals $t_{n+1}-t_n$ are small).
We refer the reader to~\cite{Hetel2017} for the survey of existing results on stability under sampled-time control.

As an alternative to periodic sampling, methods of non-uniform \emph{event-based} sampling have been proposed~\cite{Astrom:2002,Tabuada:2007}. With these methods, the next sampling instant
instant $t_{n+1}$ is triggered by some event, depending on the previous instant $t_n$ and the system's trajectory for $t>t_n$. Special cases are \emph{self-triggered} controllers~\cite{AntaTabuada:2010,MazoAntaTabuada:2010}, where $t_{n+1}$ is determined by $t_n$ and $x(t_n)$, and there is no need to check triggering conditions, and \emph{periodic} even-triggered control~\cite{HeemelsDonkers:2013}, which requires to check the triggering condition only periodically at times $n\tau$. The advantages of event-triggered control over traditional periodic control, in particular the economy of communication and energy resources, have been discussed in the recent papers~\cite{Astrom:2002,Tabuada:2007,BorgersHeemels:2014,Araujo:2014}.
Event-triggered control algorithms are widespread in biology, e.g. oscillator networks~\cite{ProCao:2017Oscill}.

A natural question arises whether a continuous-time CLF can be employed to design an \emph{event-triggered} stabilizing controller. Up to now, only a few results of this type have been reported in the literature. In~\cite{Tabuada:2007}, an event-triggered controller requires the existence of a so-called \emph{ISS Lyapunov function} $V(x)$ and a controller $u=k(x)$, satisfying the conditions
\begin{gather}
\alpha_1(|x|)\le V(x)\le \alpha_2(|x|)\quad\forall x\in\r^d\label{eq.isss}\\
V'(x)F(x,k(x+e))\le -\alpha_3(|x|)+\gamma(|e|)\quad\forall x,e\in\r^d.\label{eq.iss}
\end{gather}
Here $\alpha_i(\cdot)$ ($i=1,2,3$) are $\mathcal K_{\infty}$-functions\footnote{A function $\alpha(\cdot)$ belongs to the class $\mathcal K_{\infty}$ if it is continuous and strictly increasing with $\alpha(0)=0$ and $\lim_{s\to\infty}\alpha(s)=\infty$.} and the mappings $k(\cdot):\r^d\to\r^m$, $F(\cdot,\cdot):\r^d\times\r^m\to\r^d$, $\alpha_3^{-1}(\cdot)$ and $\gamma(\cdot):\r_+\to\r_+$ are assumed to be locally Lipschitz. Subsituting $e=0$ into~\eqref{eq.iss}, one easily shows that the ISS Lyapunov function satisfies~\eqref{eq.inf-u}, being thus a special case of CLF;
the corresponding feedback $\u(x)\dfb k(x)$ not only stabilizes the system, but in fact also provides its \emph{input to state} stability (ISS) with respect to the measurement error $e$. The event-triggered controller, offered in~\cite{Tabuada:2007}, is as follows
\be\label{eq.tabuada}
\begin{gathered}
u(t)=k(x(t_n))\quad \text{if $t\in [t_n,t_{n+1})$}\\
t_0=0,\;\; t_{n+1}=\inf\left\{t>t_n:\gamma(|e(t)|)=\sigma\alpha_3(|x(t)|)\right\},\\
e(t)=x(t_n)-x(t),\quad \sigma=const\in (0,1).
\end{gathered}
\ee
The controller~\eqref{eq.tabuada} guarantees a positive \emph{dwell time}  between consecutive events $\tau=\inf_{n\ge 0}(t_{n+1}-t_n)>0$, which is \emph{uniformly} positive for the solutions, starting in a compact set.

Whereas the condition~\eqref{eq.iss} holds for linear systems~\cite{Tabuada:2007} and some polynomial systems~\cite{AntaTabuada:2010}, in general it is restrictive and not easy to verify.
Another approach to CLF-based design of event-triggered controllers has been proposed in~\cite{Marchand:2013,Marchand:2013IFAC}.
Discarding the ISS condition~\eqref{eq.iss}, this approach is based on Sontag's theory~\cite{Sontag:1989} and inherits its basic assumptions: first, the system has to be affine
$F(x,u)=f(x)+g(x)u$, where $f,g\in C^1$, second, Sontag's controller is admissible ($u(x)\in U$ for any $x$). The controllers from~\cite{Marchand:2013,Marchand:2013IFAC} also provide positivity of the dwell time (``minimal inter-sampling interval'').

An alternative event-triggered control algorithm, substantially relaxing the ISS condition~\eqref{eq.iss} and applicable to \emph{non-affine} systems, has been proposed in~\cite{SeuretPrieurMarchand:2013} and requires the existence of a CLF, satisfying~\eqref{eq.isss} and~\eqref{eq.iss} with $e=0$
\be\label{eq.iss1}
\begin{gathered}
V'(x)F(x,k(x))\le -\alpha_3(|x|).
\end{gathered}
\ee
The events are triggered in a way providing that $V$ strictly decreases along any non-equilibrium trajectory
\small
\be\label{eq.marchand-trig}
t_{n+1}=\inf\{t\ge t_n: V'(x(t))F(x(t),u_n)=-\mu(|x(t)|)\}.
\ee\normalsize
Here $0<\mu(r)<\alpha_3(r)$ for any $r>0$ and $\mu$ is $\mathcal K_{\infty}$-function. As noticed in~\cite{SeuretPrieurMarchand:2013}, this algorithm in general \emph{does not} provide dwell time positivity, and may even lead to Zeno solutions.

As will be discussed below, the conditions~\eqref{eq.isss} and \eqref{eq.iss1} entail an estimate for the CLF's \emph{convergence rate}. In this paper, we assume that the CLF satisfies a more general convergence rate condition, and design an event-triggered controller that
preserves the convergence rate and provides positive dwell time between consecutive switchings. Also, we show that for each bounded region of the state space, self-triggered and periodic event-triggered controllers exist that provide stability for any initial condition from this region. Our approach substantially differs from the previous works~\cite{Tabuada:2007,AntaTabuada:2010,Marchand:2013,Marchand:2013IFAC,SeuretPrieurMarchand:2013}.
Unlike~\cite{Tabuada:2007,AntaTabuada:2010}, we do not assume that CLF satisfies the ISS condition~\eqref{eq.iss}. Unlike~\cite{Marchand:2013,Marchand:2013IFAC}, the affinity of the system
is not needed, and the solution's convergence rate can be explicitly estimated. Unlike~\cite{SeuretPrieurMarchand:2013}, the dwell time positivity is established.

\subsection{CLF with known convergence rate}

Whereas the existence of CLF typically allows to find a stabilizing controller, it can potentially be unsatisfactory due to very slow convergence.
Throughout the paper, we assume that a CLF gives a controller with \emph{known} convergence rate.
\begin{defn}
Consider a continuous function $\gamma:[0;\infty)\to [0;\infty)$, such that $\gamma(v)>0\,\forall v>0$ (and hence $\gamma(0)\ge 0$). A function $V(x)$, satisfying~\eqref{eq.pos-def},  is said to
be a $\gamma$-stabilizing CLF, if there exists a map $\u:\r^d\to U$, satisfying the conditions
\be\label{eq.inf-u-gamma}
V'(x)F(x,\u(x))\le -\gamma(V(x))\;\forall x,\quad F(0,\u(0))=0.
\ee
\end{defn}
\vskip 1mm
\begin{remark}\label{rem.diff1}
The condition~\eqref{eq.iss1}, as well as the stronger ISS condition~\eqref{eq.iss}, imply that $V$ is $\gamma$-stabilizing CLF with $\gamma(v)=\alpha_3\circ\alpha_2^{-1}(v)$ ($\gamma$ is continuous since $\alpha_i$ are $\mathcal K_{\infty}$-functions). In general, neither $\gamma$-CLF $V(x)$ is a monotone function of the norm $|x|$, nor $\gamma$ is monotone. Hence~\eqref{eq.inf-u-gamma}
is more general than~\eqref{eq.iss1}. Note that $\u(\cdot)$ may be \emph{discontinuous} and ``infeasible'' (the closed-loop system may have no solutions).
\end{remark}

To examine the behavior of solutions of the closed-loop system, we introduce the following function $\Gamma:(0,\infty)\to\r$
\be\label{eq.Gamma}
\Gamma(s)\dfb\int_1^s\frac{dv}{\gamma(v)},\quad s>0.
\ee
The definition~\eqref{eq.Gamma} implies that $\Gamma(s)$ is positive when $s>1$ and negative for $s<1$. Since, $\Gamma'(s)=1/\gamma(s)>0$, $\Gamma$ is increasing and hence the limits (possibly, infinite) exist
\[
\underline\Gamma\dfb\lim_{s\to 0}\Gamma(s)<0,\quad \overline\Gamma\dfb\lim_{s\to \infty}\Gamma(s)>0.
\]
The inverse $\Gamma^{-1}:(\underline\Gamma,\overline\Gamma)\to (0,\infty)$ is increasing and $C^1$-smooth. If $\underline\Gamma>-\infty$, we define
$\Gamma^{-1}(r)\dfb 0$ for $r\le\underline\Gamma$.

To understand the meaning of the function $\Gamma(s)$, consider now a special situation, where the equality in~\eqref{eq.inf-u-gamma} is achieved
\be\label{eq.inf-u-gamma-aux}
V'(x)F(x,\u(x))= -\gamma(V(x))\;\forall x\in\r^d.
\ee
The CLF $V(x(t))$ can be treated as some ``energy'', stored in the system at time $t$, whereas $\gamma(V(x(t)))=-\dot V(x(t))$ can be treated as the energy dissipation rate or ``power''
consumed by the closed-loop system (``work'' done by the system per unit of time) with feedback $u=\u(x)$. By noticing that
$\frac{d}{dt}\Gamma(V(x(t))=\dot V(x(t))/\gamma(V(x(t))=-1$, the function $\Gamma$ may be considered as the ``energy-time characteristics'' of the system:
it takes the system time $t_1=\Gamma(V_0)-\Gamma(V_1)$ to move from the energy level $V_0=V(x(0))$ to the energy level $V_1$.

In general,~\eqref{eq.inf-u-gamma} implies an upper bound for a solution.
\begin{prop}\label{prop.converge}
Let the system~\eqref{eq.syst} have a $\gamma$-stabilizing CLF $V$, corresponding to the controller $\u$. Let $x(t)$ be a solution to
\be\label{eq.closed-loop}
\dot x(t)=F(x(t),u(t)),\quad u(t)\dfb \u(x(t)).
\ee
Then on the interval of the solution's existence the function $V(t)=V(x(t))$ satisfies the following inequality
\be\label{eq.conv-rate}
0\le V(t)\le \Gamma^{-1}(\Gamma(V(0))-t).
\ee
\end{prop}
\begin{proof}
If $V(t)>0$ at any time when the solution exists, then $\dot V(t)=V'(x(t))F(x(t),u(t))\overset{\eqref{eq.inf-u-gamma}}{\le} -\gamma(V(t))<0$ and
\be\label{eq.aux1}
\frac{d}{dt}\Gamma(V(t))\le -1\Longrightarrow \Gamma(V(t))\le\Gamma(V(0))-t,
\ee
which implies~\eqref{eq.conv-rate} since $\Gamma^{-1}$ is increasing. Suppose now that $V(t)$ vanishes at some $t\in [0,\delta)$, and let $t_0\ge 0$ be the first such instant.
By definition, for $t\in [0,t_0)$ one has $V(t)>0$, which entails~\eqref{eq.aux1} and~\eqref{eq.conv-rate}. Since $V$ is non-increasing, $V(t)\equiv 0$ for $t\ge t_0$, and thus~\eqref{eq.conv-rate} holds also for $t\ge t_0$.
\end{proof}
\begin{corollary}
If $\underline\Gamma>-\infty$, then the solution of~\eqref{eq.closed-loop} converges to $0$ in \emph{finite time} $\delta_*=\Gamma\left(V(x(0))\right)-\underline\Gamma$ (provided that it exists on $[0,\delta_*)$. If $\underline\Gamma=-\infty$ and $x(t)$ is a forward complete solution to~\eqref{eq.closed-loop}, then $x(t)\xrightarrow[t\to\infty]{} 0$.
\end{corollary}

Depending on the finiteness of $\underline\Gamma$, Proposition~\eqref{prop.converge} explicitly estimates either \emph{time} or \emph{rate} of the CLF's convergence to $0$.

\textbf{Example 1.} Let $\gamma(v)=\ae v$, where $\ae>0$ is a constant. In this case $\Gamma(s)=\ae^{-1}\ln s$, $\underline\Gamma=-\infty$, $\overline\Gamma=\infty$, $\Gamma^{-1}(r)=e^{\ae r}$. The $\gamma$-stabilizing CLF provides \emph{exponential} stabilization (being an ES-CLF~\cite{Ames:2014}). The inequality~\eqref{eq.conv-rate} reduces to
\be\label{eq.es-clf}
0\le V(t)\le \exp\left(\ae(\ae^{-1}\ln V(0)-t)\right)=V(0)e^{-\ae t}.
\ee

\textbf{Example 2.}  Let $\gamma(v)=\ae v^{a}$ with $\ae>0,a>1$. We have $\Gamma(s)=[\ae(a-1)]^{-1}(1-s^{1-a})$, $\underline\Gamma=-\infty$, $\overline\Gamma=[\ae(a-1)]^{-1}$, $\Gamma^{-1}(r)=\left(1-\ae(a-1)r\right)^{1/(1-a)}$, and~\eqref{eq.conv-rate} boils down to
\be\label{eq.aux2}
\begin{aligned}
V(t)\leq 
\left(V(0)^{1-a}+t\ae(a-1)\right)^{\frac{1}{1-a}}.
\end{aligned}
\ee

\textbf{Example 3.}  Let $\gamma(v)=\ae v^{a}$ with $\ae>0,a<1$. Similar to the case $a>1$, one has $\Gamma(s)=[\ae(a-1)]^{-1}(1-s^{1-a})$ and $\Gamma^{-1}(r)=\left(1-\ae(a-1)r\right)^{1/(1-a)}$, however, $\underline\Gamma=[\ae(a-1)]^{-1}>-\infty$. The condition~\eqref{eq.conv-rate} again leads to~\eqref{eq.aux2}, however, the right-hand side
vanishes for $t\ge t_0\dfb \vk(1-a)^{-1}V(0)^{1-a}$, e.g. the solution converges in finite time $t_0$.

Example~3 shows that a CLF can give a controller, solving the problem of \emph{finite-time} stabilization. An event-triggered counterpart of such a controller can be designed, using the procedure discussed in the next section.
However, the property of local positivity of dwell time does not hold for such a controller (see Remark~\ref{rem.dwell-time}), and thus the absence of Zeno trajectories does not follow from our main results. Finite-time event-triggered stabilization is thus beyond the scope of this paper, being a subject of ongoing research.

\subsection{Problem setup}

In this paper, we address the following fundamental question: does the existence of a continuous-time $\gamma$-stabilizing CLF allow to design an \emph{event-triggered} mechanism, providing the same convergence rate as the continuous-time control $u=\u(x)$? Relaxing the latter requirement, we seek for event-triggered controllers whose convergence rates are \emph{arbitrarily close} to the
convergence rate of the continuous-time controller.

\textbf{Problem.} Assume that $V$ is a $\gamma$-stabilizing CLF, where $\gamma(v)$ is a known function, and $\sigma\in (0,1)$ is a fixed constant.
Design an event-triggered controller, providing the following condition
\be\label{eq.inf-u-sigma0}
\dot{V}(x(t))\le-\sigma\gamma(V(x(t)))\quad\forall t\ge 0.
\ee

Applying Proposition~\ref{prop.converge} to $\tilde \gamma(s)=\sigma\gamma(s)$ (which corresponds to $\tilde\Gamma(s)=\sigma^{-1}\Gamma(s)$),
it is shown that~\eqref{eq.inf-u-sigma0} entails that
\be\label{eq.conv-rate1}
0\le V(x(t))\le \Gamma^{-1}(\Gamma(V(x(0)))-\sigma t).
\ee
For instance, in the Example~1 considered above \eqref{eq.conv-rate1} implies exponential convergence with exponent $\sigma\ae$ (that is, $V(t)\le V(0)e^{-\sigma\ae t}$) (versus the rate $\ae$ in
continuous time).

\begin{remark}\label{rem.cbf}
In some situations, the CLF serves not only as a Lyapunov function, but also as a \emph{barrier certificate}~\cite{RomdlonyJayawardhana:2016}, ensuring that the trajectories do not cross some ``unsafe'' set $\mathcal D$. For instance, suppose that for any point of the boundary $\xi\in\partial\mathcal D$ one has $V(\xi)\ge v_*>0$. Then for any initial condition beyond the unsafe set's closure $x(0)\not\in \overline{\mathcal D}$ such that $V(x(0))<v_*$, the solution of the continuous-time system~\eqref{eq.closed-loop} starting at $x(0)$ cannot cross the boundary $\partial\mathcal D$ and thus cannot enter the unsafe set. The event-triggered algorithm providing~\eqref{eq.inf-u-sigma0} preserves the latter property of the CLF and provides thus safety for the aforementioned class of initial conditions.
\end{remark}

\section{Event-triggered, Self-Triggered and Periodic Event-Triggered Controller Designs}\label{sec.event}

Henceforth we suppose that a continuous strictly positive function $\gamma(\cdot)$, a $\gamma$-stabilizing CLF $V(x)$ and the corresponding feedback map $\u:\r^d\to U$ are fixed.
All algorithms considered in this paper provide that $u(t)\in \u(\r^d)$; without loss of generality, we assume that $U=\u(\r^d)$. We are going to design an event-triggered algorithm that ensures~\eqref{eq.inf-u-sigma0}. The input $u(t)$ switches at sampling instants $t_0,t_1,\ldots$, where $t_0=0$ and the next instants $t_n$ depends on the solution, remaining constant $u(t)\equiv u_n=u(t_n)$ on each sampling interval $[t_n,t_{n+1})$.

\subsection{The event-triggered control algorithm design}

The condition~\eqref{eq.inf-u-sigma0} can be rewritten as $W(x(t),u(t))\leq -\sigma\gamma(V(x(t))$, where the function $W$ is defined by
\be\label{eq.w}
\begin{gathered}
W(x,u)\dfb V'(x)F(x,u)\in\r,\quad x\in\r^d, u\in U,
\end{gathered}
\ee

At the initial instant $t_0=0$, calculate the control input $u_0\dfb\u(x(t_0))$.
If $V(x(t_0))=0$, then the system starts at the equilibrium point and stays there under the control input $u(t)\equiv u_0\quad\forall t\ge t_0$.
Otherwise, $W(x(t_0),u(t_0))\leq -\gamma(V(x(t_0)))<-\sigma\gamma(V(x(t_0)))$ due to~\eqref{eq.inf-u-gamma}, and hence for $t$ sufficiently close to $t_0$ one has
$
W(x(t),u_0)<-\sigma\gamma(V(x(t))).
$
The next sampling instant $t_1$ is the \emph{first} time when
\ben
W(x(t),u_0)=-\sigma\gamma(V(x(t))),
\een
we formally define $t_1=\infty$ if such an instant does not exist. If $t_1<\infty$, we repeat the procedure, calculating the new control input $u_1=\u(x(t_1))$.
If $V(x(t_1))=0$, then the system has arrived at the equilibrium, and stays there under the control input $u(t)\equiv u_1$. Otherwise, $W(x(t_1),u(t_1))\overset{\eqref{eq.inf-u-gamma}}{\le} -\gamma(V(x(t_1)))<-\sigma\gamma(V(x(t_1)))$. Hence for $t$ close to $t_1$ one has
$
W(x(t),u_1)<-\sigma\gamma(V(x(t))).
$
The next sampling instant $t_2$ is the first time $t>t_1$ when $W(x(t),u_1)=-\sigma\gamma(V(x(t)))$, we define $t_2=\infty$ if such an instant does not exist. Iterating this procedure,
the sequence of instants sampling $t_0<t_1<\ldots<t_n<t_{n+1}<\ldots$ is constructed in a way that the control $u(t)=u_n\dfb\u(x(t_n))$ for $t\in[t_n,t_{n+1})$ satisfies~\eqref{eq.inf-u-gamma}.
If $V(x(t_n))>0$, $t_{n+1}$ is the first time $t>t_n$ when
\be\label{eq.inf-u-sigma-eq-n}
W(x(t),u_n)=-\sigma\gamma(V(x(t))).
\ee
The sequence of sampling instants terminates if $V(x(t_n))=0$ or~\eqref{eq.inf-u-sigma-eq-n} does not hold at any $t>t_n$, in this case we formally define $t_{n+1}=\infty$
and the control is frozen $u(t)\equiv u_n\,\forall t>t_n$.

The procedure just described can be written as follows
\be\label{eq.alg1}
\begin{gathered}
u(t)=\u(x(t_n))\;\;\forall t\in [t_n,t_{n+1}),\quad t_0=0,\\
t_{n+1}=\begin{cases}\inf\left\{t>t_n: \eqref{eq.inf-u-sigma-eq-n}\;\text{holds} \right\},\,&V(x(t_n))> 0,\\
\infty,\,&V(x(t_n))=0.
\end{cases}
\end{gathered}
\ee
(where $\inf\emptyset=+\infty$), or in the following ``pseudocode form''.
\begin{algorithm}[H]
\begin{algorithmic}
\State $n\gets 0$; $t_n\gets 0$; $u_n\gets\u(x(0))$;
\While{$V(x(t_n))>0$}
\Repeat
    \State $u(t)=u_n$; \Comment{$t$ is the current time}
\Until{$W(x(t),u_n)=-\sigma \gamma(V(x(t)))$};
     \State $n\gets n+1$; $t_n\gets t$; $u_n\gets\u(x(t_n)))$;
\EndWhile ;
\State \textbf{freeze $u(t)\equiv \u(0)$}; \Comment{stay in the equilibrium}
\end{algorithmic}
\caption*{Algorithm~\eqref{eq.alg1} in the pseudocode form}\label{alg.1code}
\end{algorithm}
\begin{remark}
Implementation of Algorithm~\eqref{eq.alg1} does not require any closed-form analytic expression for $\u(x)$; if suffices to have some numerical algorithm for computation of the value $u_n=\u(x(t_n))$ at a specific point $x(t_n)$.
\end{remark}
\begin{remark}
Triggering condition~\eqref{eq.inf-u-sigma-eq-n} is similar to the condition~\eqref{eq.marchand-trig}, employed by the algorithm from~\cite{SeuretPrieurMarchand:2013},
however, as explained in Remark~\ref{rem.diff1}, in general the conditions adopted in~\cite{SeuretPrieurMarchand:2013} do not hold. Furthermore,
unlike~\cite{SeuretPrieurMarchand:2013}, we give conditions for the positivity of dwell time (to be defined below) and explicitly estimate the convergence rate of the algorithm.
\end{remark}

To assure the practical applicability of the algorithm~\eqref{eq.alg1}, one has to prove that the solution of the closed-loop system is forward complete, addressing thus two problems. The first problem, addressed in Subsection~\ref{sec.event}-B, is the solution existence between two sampling instants: to show that the event~\eqref{eq.inf-u-sigma-eq-n} is detected earlier than the solution to the following equation ``explodes'' (escapes from any compact)
\be\label{eq.cauchy-n}
\dot x(t)=F(x(t),u_n),\quad u_n=\u(x(t_n)),\,t\ge t_n.
\ee
The second problem, addressed in Subsection~\ref{sec.event}-C, is to show the impossibility of Zeno solutions.
\begin{defn}
A solution to the closed-loop system~\eqref{eq.syst},\eqref{eq.alg1} is said to be \emph{Zeno}, or exhibit \emph{Zeno behavior}
if the sequence of sampling instants is infinite and has a limit $t_{\infty}=\lim\limits_{n\to\infty}t_n=\sup\limits_{n\ge 0}t_n<\infty$; otherwise, the trajectory is said to be \emph{non-Zeno}.
\end{defn}

Although mathematically it can be possible to prolong the solution beyond the time $t_{\infty}$~\cite{AmesZheng:2006}, the practical implementation of algorithm~\eqref{eq.alg1} with Zeno trajectories is problematic. Moreover, any real-time implementation of the algorithm imposes an implicit restriction on the minimal time between two consecutive events, referred to as the solution's \emph{dwell-time}. Since the control commands cannot be computed arbitrarily fast, in practice the solutions with zero dwell-time are also undesirable, even if they are forward complete.
\begin{defn}
The value $\mathfrak{T}(x_0)=\inf\limits_{n\ge 0}(t_{n+1}(x_0)-t_n(x_0))$ is called the \emph{dwell-time} or the minimal inter-sampling interval (MSI)~\cite{Marchand:2013} of the solution.
Algorithm~\eqref{eq.alg1} provides \emph{locally uniformly} positive dwell-time if $\mathfrak{T}$ is uniformly positive over all solutions, starting in a compact set $\mathcal{K}$: $\inf\limits_{x_0\in \mathcal K}\mathfrak{T}(x_0)>0$.
\end{defn}

The proof of locally uniform dwell-time positivity allows to design self-triggered and periodic event-triggered modifications of~\eqref{eq.alg1} that are discussed in Subsections~\ref{sec.event}-D,E.

\begin{remark}\label{rem.dwell-time}
By definition of the dwell-time, $t_1-t_0=t_1\ge \mathfrak{T}(x(0))$. In particular, if $\u(x(0))\ne \u(0)$, then $x(t)\ne 0$ for $t\in [0,\mathfrak{T}(x(0))$ (when $x=0$, the control has to be switched to $\u(0)$).
For instance, in the situation from Example~3 from previous section, the solution (if it exists) converges to $0$ in time, proportional to $V(x(0))$ due to~\eqref{eq.conv-rate1}. Such a controller can provide the dwell-time positivity, but
not locally uniform positivity since $\mathfrak{T}(x_0)\le\sigma^{-1}V(x_0)\to 0$ as $|x_0|\to 0$.
\end{remark}

\textcolor{black}{Remark~\ref{rem.dwell-time} may be illustrated by the simple example of the system $\dot x=u$ and a relay control $\u(x)={\rm sgn}\,x$. Choosing $V(x)=x^2$ and $\gamma(v)=2\sqrt{v}$, the event-triggered algorithm~\eqref{eq.alg1} in fact coincides with the continuous time control: the first event is fired at time $t_0$ and $u_0={\rm sgn}\,x_0$; if $x_0\ne 0$, the second event occurs at $t_1=|x_0|$ and $u_1=0$.
}

\color{black}

\subsection{The inter-sampling behavior of solutions}

To examine the solutions' behavior between two sampling instants, we introduce the auxiliary Cauchy problem
\be\label{eq.cauchy}
\dot \xi(t)=F(\xi(t),u_*),\; \xi(0)=\xi_0,\;t\ge 0,
\ee
where $u_*\in U$. To provide the unique solvability of~\eqref{eq.cauchy}, henceforth the following non-restrictive assumption is adopted.
\begin{assum}\label{ass.contin}
For $u_*\in U$, the map $F(\cdot,u_*)$ is locally Lipschitz; in particular, $W(\cdot,u_*):\r^d\to\r$ is continuous\footnote{Recall that $V\in C^1$ by definition of the CLF}.
\end{assum}

\begin{prop}\label{prop.tech}
Under Assumption~\ref{ass.contin}, the Cauchy problem~\eqref{eq.cauchy} has the unique solution $\xi(t)=\xi(t|\xi_0,u_*)$, which satisfies at least one of the following two conditions holds
\begin{enumerate}
\item $W(\xi(t),u_*)>-\sigma\gamma(V(\xi(t)))$ for some $t\ge 0$;
\item the solution is bounded and forward complete.
\end{enumerate}
\end{prop}
\begin{proof}
The first statement follows from the Picard-Lindel\"of existence theorem~\cite{Khalil}. Assume that on the interval of the solution's existence we have
$\dot{V}(\xi(t))=W(\xi(t),u_*)\leq -\sigma\gamma(V(\xi(t)))$ (the first condition does not hold). Then $V(\xi(t))\le V(\xi_0)$, and hence $\xi(t)$ also remains bounded on its interval of existence,
and hence is forward complete.
\end{proof}
\begin{corollary}\label{cor.unique1}
Under Assumption~\ref{ass.contin}, $x(t)=\xi(t-t_+|x_+,u_*)$ is the only solution to the following Cauchy problem
\be\label{eq.cauchy+}
\dot x(t)=F(x(t),u_*),\; x(t_+)=x_+,\;t\ge t_+,
\ee
where $u_*\in U$. If $x_+=0$ and $u_*=\u(0)$, then $\xi(t)\equiv 0$.
\end{corollary}

Corollary~\ref{cor.unique1} allows to show that the solution to the closed-loop system~\eqref{eq.syst},\eqref{eq.alg1} exists and unique for any initial condition.
One can show via induction on $n$ that the sequence $\{t_n\}$ is uniquely defined by $x(0)$ by noticing that
$t_0=0$ is uniquely defined and if $t_n<\infty$, then the next instant $t_{n+1}\leq\infty$ depends only on $t_n,x_n,u_n$. If $x_n=0$, then algorithms stops and $t_{n+1}=\infty$. In view of Proposition~\ref{prop.tech}, either event~\eqref{eq.inf-u-sigma-eq-n} occurs at some time $t>t_n$ (the first such instant is $t_{n+1}<\infty$), or the solution is well defined on $[t_n,\infty)$ and satisfies~\eqref{eq.inf-u-sigma0} (in which case $t_{n+1}=\infty$). In both situations, the solution is well defined on the $n$th sampling interval $[t_n,t_{n+1})$.
\begin{corollary}\label{cor.t-n-exist}
Let Assumption~\ref{ass.contin} hold. Then the sequence of sampling instants $\{t_n\}$ in the algorithm~\eqref{eq.alg1}
is uniquely defined by the initial condition $x(0)$, and the solution between them is uniquely defined by the formula
\be\label{eq.aux0}
x(t)=\xi(t-t_n|x_n,u_n)\quad\forall t\in [t_n,t_{n+1}).
\ee
where $\xi(t|\xi_0,u_*)$ stands for the solution to~\eqref{eq.cauchy}.
\end{corollary}

Notice that the solution is automatically forward complete in the case where the sequence $t_n$ terminates (for some $n$, we have $t_{n+1}=\infty$).
This however is \emph{not} guaranteed for the case where infinitely many events occur. To exclude the possibility of Zeno behavior, additional assumptions are required.

\subsection{Dwell time positivity}

In this subsection, we formulate our first main result, namely, the criterion of dwell time positivity in Algorithm~\eqref{eq.alg1}.
This criterion relies on several additional assumptions.

For any $x_*\in\r^d$ and $\mathcal K\subset\r^d$, denote
\be\label{eq.b}
B(x_*)\dfb\{x: V(x)\le V(x_*)\},\;\; B(\mathcal K)\dfb \bigcup\limits_{x_*\in \mathcal K}B(x_*).
\ee
Algorithm~\eqref{eq.alg1} implies that $V(x(t))$ is non-increasing due to~\eqref{eq.inf-u-gamma}, and hence $x(t)\in B(x(s))$ for $t\ge s\ge 0$.
In particular, sets $B(x_*)$ are \emph{forward invariant} along the solutions of~\eqref{eq.syst},\eqref{eq.alg1}.
For any bounded set $\mathcal K$, $B(\mathcal K)$ is also bounded since
\[
B(\mathcal K)\subseteq\{x: V(x)\le \sup_{x_*\in \mathcal K} V(x_*)\}.
\]
Accordingly to Assumption~\ref{ass.contin}, the following supremum is finite
\be\label{eq.kappa}
\vk(x_*)\dfb\sup\limits_{\substack{x_1,x_2\in B(x_*)\\x_1\ne x_2}} \frac{|F(x_1,\u(x_*))-F(x_2,\u(x_*))|}{|x_2-x_1|}<\infty\\
\ee
for any $x_*$ (in the case where $x_*=0$ and $B(x_*)=\{0\}$, let $\vk(x_*)\dfb 0$). We adopt a stronger version of Assumption~\ref{ass.contin}.
\begin{assum}\label{ass.F}
The Lipschitz constant $\vk(x_*)$ in~\eqref{eq.kappa} is a \emph{locally bounded} function of $x_*$.
\end{assum}

Assumption~\ref{ass.F} holds, for instance, if the mapping $\u$ is locally bounded and $F'_x(x,u)$ exists and is continuous in $x$ and $u$.

\begin{assum}\label{ass.gradient}
The gradient $V'(x)$ is locally Lipschitz.
\end{assum}

Assumption~\ref{ass.gradient} is a stronger version of CLF's smoothness and holds e.g. when $V\in C^2$. Similar to~\eqref{eq.kappa}, we
introduce the Lipschitz constant of $V'$ on the compact set $B(x_*)$:
\be\label{eq.nu}
\nu(x_*)\dfb\sup\limits_{\substack{x_1,x_2\in B(x_*)\\x_1\ne x_2}} \frac{|V'(x_1)-V'(x_2)|}{|x_2-x_1|},\quad\nu(0)\dfb 0.
\ee

Assumption~\ref{ass.gradient} implies that $\nu$ is \emph{locally bounded} since for any compact $\mathcal{K}$ the set $B(\mathcal K)$ is bounded and
\[
\sup_{x_*\in \mathcal K}\nu(x_*)\le \sup\limits_{\substack{x_1,x_2\in B(\mathcal K)\\x_1\ne x_2}} \frac{|V'(x_1)-V'(x_2)|}{|x_2-x_1|}<\infty.
\]

\begin{figure}[h]
\center
\includegraphics[height=3cm]{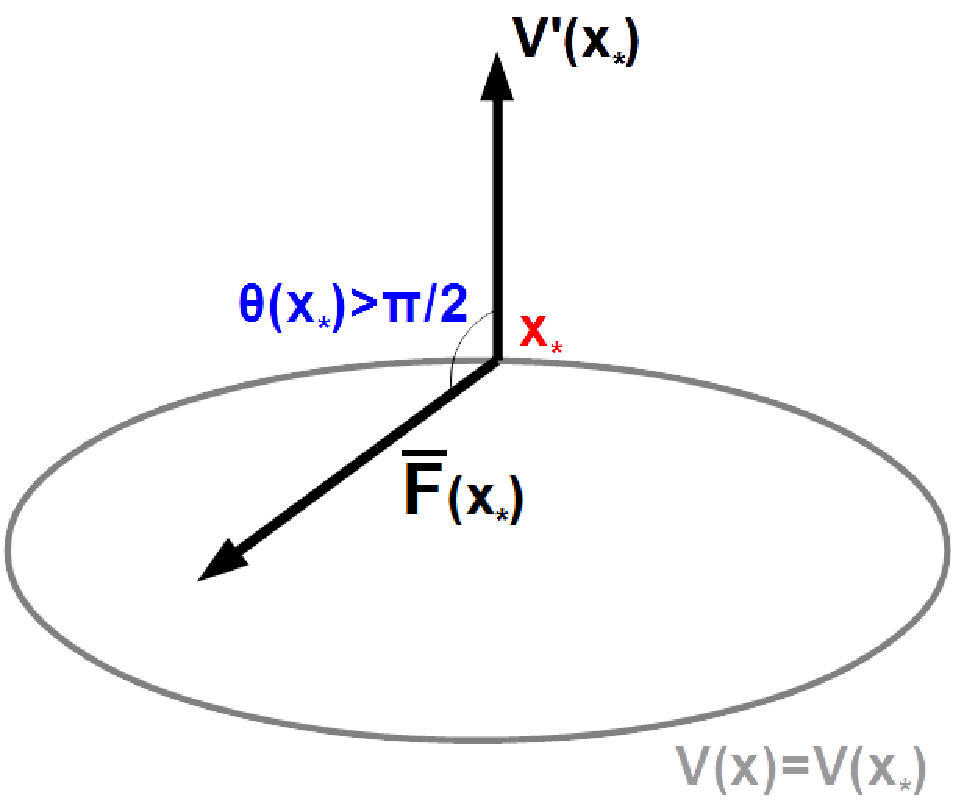}
\caption{Illustration to Assumption~\ref{ass.non-degen}: the angle $\theta(x_*)$}\label{fig.1}
\end{figure}

Finally, we adopt an assumption that allows to establish the relation between the convergence rates of the $\gamma$-CLF $V(x(t))$ under the continuous-time control $\u=\u(x)$ and the solution $x(t)$. Notice that~\eqref{eq.inf-u-gamma} gives no information about the speed of the solution's convergence since $\dot V(x)=V'(x)\dot x(t)$ depends only on the velocity's $\dot x(t)$ projection on the gradient vector $V'(x)$, whereas its transversal component can be arbitrary. These transversal dynamics can potentially lead to very slow and ``non-smooth'' convergence, 
\textcolor{black}{in the sense that $|\dot x(t)|\gg |\dot V(x(t))|$. As discussed in Appendix~\ref{app.discuss}, in such a situation the dwell-time positivity cannot be proved}. Denoting
\[
\bar F(x)\dfb F(x,\u(x)),
\]
\textcolor{black}{and introducing the angle $\theta(x)$ between $\bar F(x)$ and $V'(x)$ (Fig.~\ref{fig.1}),}
the definition of $\gamma$-CLF~\eqref{eq.inf-u-gamma} implies that
\[
\begin{gathered}
V'(x)=0\Longrightarrow x=0\Longrightarrow \bar F(x)=0\\
\cos\theta(x)<0\quad\forall x\ne 0.
\end{gathered}
\]
Our final assumption requires these conditions to hold uniformly in the vicinity of $x=0$ in the following sense.
\begin{assum}\label{ass.non-degen}
The $\gamma$-CLF $V(x)$ and the corresponding controller $\u(x)$ satisfy the following properties:
\be\label{eq.non-degen}
\begin{gathered}
|\bar F(x)|\le M_1(x)|V'(x)|\quad\forall x\in\r^d,\\
\cos\theta(x)\le -M_2(x)\quad\forall x\in\r^d\setminus\{0\},
\end{gathered}
\ee
where the functions $M_1,M_2$ are, respectively, uniformly bounded and uniformly strictly positive on any compact set.
\end{assum}

The inequalities~\eqref{eq.non-degen} imply that the solution does not oscillate near the equilibrium since $|\bar F(x)|\to 0$ as $|x|\to 0$, and the angle between the vectors\footnote{The inequality~\eqref{eq.inf-u-gamma} implies that both vectors are non-zero unless $x\ne 0$.} $\dot x=\bar F(x)$ and $V'(x)$ remains strictly obtuse as $x\to 0$, i.e. the flow is not transversal to the CLF's gradient. Assumption~\ref{ass.non-degen} can be reformulated as follows.
\begin{lemma}\label{lem.tech}
For a $\gamma$-CLF $V$, Assumption~\ref{ass.non-degen} holds if and only if
a locally bounded function $M(x)>0$ exists such that
\be\label{eq.non-degen1}
|V'(x)|\,|\bar F(x)|+|\bar F(x)|^2\leq M(x)|V'(x)\bar F(x)|\,\forall x\in\r^d.
\ee
\end{lemma}
\begin{proof}
\textcolor{black}{For $M(x)\dfb (1+M_1(x))/M_2(x)$,~\eqref{eq.non-degen} implies
\[
\begin{split}
M(x)|V'(x)\bar F(x)|=M(x)|\cos\theta(x)||V'(x)|\,|\bar F(x)|\\ \overset{\eqref{eq.non-degen}}{\geq} M(x)M_2(x)|V'(x)|\,|\bar F(x)|\overset{MM_2=1+M_1}{\geq} \\
\geq |V'(x)|\,|\bar F(x)|+M_1(x)|V'(x)|\,|\bar F(x)|\\
\overset{\eqref{eq.non-degen}}{\geq} |V'(x)|\,|\bar F(x)|+|\bar F(x)|^2,
\end{split}
\]
proving thus the ``only if'' part.} To prove the ``if'' part, note that~\eqref{eq.non-degen1} and~\eqref{eq.inf-u-gamma} imply the inequalities
\[
\begin{gathered}
M(x)\cos\theta(x)=\frac{M(x)V'(x)\bar F(x)}{|V'(x)|\,|\bar F(x)|}\leq -1\\
|\bar F(x)|^2\leq M(x)|V'(x)\bar F(x)|\le M(x)|V'(x)|\,|\bar F(x)|,
\end{gathered}
\]
and hence~\eqref{eq.non-degen} holds with $M_1=M$ and $M_2=1/M$.
\end{proof}

We not turn to the key problem of dwell time estimation for Algorithm~\eqref{eq.alg1}. In view of~\eqref{eq.aux0}, to estimate of the time elapsed between consecutive events $t_{n+1}-t_n$, it suffices to study the behavior of the solution $\xi(t)=\xi(t|x_*,\u(x_*))$ to the Cauchy problem~\eqref{eq.cauchy} with $\xi_0=x_*\ne 0$ and $u_*=\u(x_*)$, namely, to find the first instant $\bar t$ such that $W(\xi(\bar t),u_*)=-\sigma\gamma(V(\xi(\bar t)))$. The following lemma implies that $\bar t\ge\tau(x_*)$, 
\textcolor{black}{where $\tau(\cdot)$ is some function, \emph{uniformly strictly positive} on any compact set.}
\begin{lemma}\label{lem.key-lemma}
Let Assumptions~\ref{ass.contin}-\ref{ass.non-degen} hold and $\gamma(\cdot)$ be either non-decreasing or $C^1$. Then a function $\tau:\r^d\to (0,\infty)$ exists, depending on $\sigma,\gamma,\vk,\nu,M$, that satisfies two conditions:
\begin{enumerate}
\item $\tau(\cdot)$ is uniformly strictly positive on any compact set;
\item for any $x_*\ne 0$, the solution $\xi(t)=\xi(t|x_*,\u(x_*))$ is well-defined on the closed interval $[0,\tau(x_*)]$ and
\be\label{eq.w-ineq-main}
W(\xi(t),\u(x_*))<-\sigma \gamma(V(\xi(t)))\;\forall t\in [0,\tau(x_*)).
\ee
\end{enumerate}
Moreover, if the functions $\vk$, $\nu$, $M$ are \emph{globally} bounded, $\gamma\in C^1$ and $\inf\limits_{v\ge 0}\gamma'(v)>-\infty$, then $\inf_{x_*\in\r^d}\tau(x_*)>0$.
\end{lemma}

The proof of Lemma~\ref{lem.key-lemma} will be given in Appendix~\ref{app.proof}; in this proof the exact expression for $\tau(\cdot)$ will be found, which
involves the functions $\gamma,\vk,\nu,M$. Note that Algorithm~\eqref{eq.alg1} \emph{does not} employ $\tau(\cdot)$, which is needed to estimate the dwell time.
Notice that for a fixed $x_*\in\r^d$, the value $\tau(x_*)=\tau_{\sigma}(x_*)$ may be considered as a function of the parameter $\sigma$ from~\eqref{eq.inf-u-sigma0}.
It can be shown that $\tau_{\sigma}(x_*)\to 0$ as $\sigma\to 1$. In other words, if the event-triggered algorithm provides the same convergence rate as the continuous-time control,
the dwell time between consecutive events vanishes. Lemma~\ref{lem.key-lemma} implies our main result.
\begin{theorem}\label{thm.dwell}
Let the assumptions of Lemma~\ref{lem.key-lemma} hold. Then the following estimate for the dwell-time in~\eqref{eq.alg1} holds
\be\label{eq.tau-min}
\mathfrak{T}(x_0)\ge \tau_{\min}(x_0)\dfb\inf_{x\in B(x_0)}\tau(x)>0,
\ee
where $\tau(x)$ stands for the function from Lemma~\ref{lem.key-lemma}. The dwell-time $\mathfrak{T}$ is \emph{uniformly positive} on any compact.
Moreover, if the functions $\vk$, $\nu$, $M$ are \emph{globally} bounded, $\gamma\in C^1$ and $\inf\limits_{v\ge 0}\gamma'(v)>-\infty$, then $\mathfrak{T}$ is uniformly strictly positive on $\r^d$.
\end{theorem}
\begin{proof}
Notice first that the function $\tau_{\min}$ from~\eqref{eq.tau-min} is locally uniformly positive on any compact set $K\subseteq\r^d$ since
\[
\inf_{x_0\in\mathcal K}\tau_{min}(x_0)=\inf_{x\in B(\mathcal K)}\tau(x)>0
\]
due to the boundedness of the set $B(\mathcal K)$ and local uniform positivity of $\tau$.
Applying Lemma~\ref{lem.key-lemma} to $x_*=x_n$ and using~\eqref{eq.aux0}, one shows that if the
$n$th event is raised at the instant $t_n<\infty$, the next event cannot be fired earlier than
at time $t_n+\tau(x_n)$. Since $x_n\in B(x_0)$, one has $t_{n+1}\ge t_n+\tau_{min}(x_0)$, which implies~\eqref{eq.tau-min} by definition of the dwell time $\mathfrak{T}(x_0)$.
\end{proof}

\subsection{Self-triggered and time-triggered stabilizing control}

As has been already mentioned, \textcolor{black}{Algorithm~\eqref{eq.alg1} requires neither full knowledge of the functions $\vk,\nu,M$, nor even upper estimates for them}. If such estimates are known, $\tau(\cdot)$ from Lemma~\ref{lem.key-lemma} can be found explicitly
(see Appendix~\ref{app.proof}), and algorithm~\eqref{eq.alg1} can be replaced by the \emph{self-triggered} controller:
\be\label{eq.alg2}
\begin{gathered}
u(t)=\u(x(t_n)),\quad t\in [t_n,t_{n+1}),\\
\quad t_0=0,\quad t_{n+1}=\begin{cases} t_n+\tau(x(t_n)),\,& V(x(t_n))>0\\ \infty, & V(x(t_n))=0.\end{cases}
\end{gathered}
\ee
The algorithm~\eqref{eq.alg2} requires to compute the value of $\tau(x_n)$ at each step. Alternatively, if a lower bound $\tau_*$ for the value of $\tau_{min}(x_0)$ from~\eqref{eq.tau-min}
is known $\tau_{min}(x_0)\geq\tau_*>0$, one may consider periodic or aperiodic \emph{time-triggered} sampling
\be\label{eq.alg2+}
\begin{gathered}
u(t)=\u(x(t_n)),\quad t\in [t_n,t_{n+1}),\\
t_0=0,\quad 0<t_{n+1}-t_n\le\tau_*,\quad \lim\limits_{n\to\infty} t_n=\infty.
\end{gathered}
\ee
Here the sequence $\{t_n\}$ is independent of the trajectory; often $t_n=n\tau_0$ with some period $\tau_0\le\tau_*$.

\begin{remark}
Notice that to find a lower estimate for $\tau_{min}(x_0)$, there is no need to know the initial condition $x_0$ (which can be uncertain); it suffices to know an \emph{upper bound} for the value of $V(x_0)$, which determines the set $B(x_0)$.
\end{remark}

Lemma~\ref{lem.key-lemma} and~\eqref{eq.aux0} yield in the following result.
\begin{theorem}\label{thm.self}
 Under the assumptions of Lemma~\ref{lem.key-lemma}, any solution to the closed-loop system~\eqref{eq.syst},\,\eqref{eq.alg2} is forward complete and satisfies~\eqref{eq.inf-u-sigma0}. The same holds for solutions to~\eqref{eq.syst},\,\eqref{eq.alg2+}, whose initial conditions satisfy the inequality $\tau_{min}(x(0))\ge\tau_*$.
\end{theorem}
\begin{proof}
Theorem~\ref{thm.self} is proved very similar to Theorem~1, with the only technical difference that~\eqref{eq.inf-u-sigma0} is not automatically guaranteed along the trajectories, and thus forward invariance of the set $B(x_0)$ still has to be proved. Using induction on $n=0,1\,\ldots,$ we are going to prove that $x(t_n)\in B(x(0))$ for each $n$. The induction base $n=0$ is obvious.
Assuming that $x(t_n)\in B(x(0))$, we know that $t_{n+1}-t_n\le \tau(x(t_n))$ (in the case of~\eqref{eq.alg2+} this holds since
$\tau_{min}(x_0)\le\tau(x(t_n))$). Substituting $x_*=x_n$ to~\eqref{eq.w-ineq-main} and using~\eqref{eq.aux0}, one shows that~\eqref{eq.inf-u-sigma0} holds on each sampling interval $[t_n,t_{n+1}]$,
and thus $x(t_{n+1})\in B(x(t_n))$. This proves the induction step, entailing also that both algorithms ensure~\eqref{eq.inf-u-sigma0}.
The solution thus remains bounded and is forward complete ($t_n\to\infty$).
\end{proof}
\begin{remark}
As follows from Lemma~\ref{lem.key-lemma}, if the functions $\vk$, $\nu$, $M$ are \emph{globally} bounded, $\gamma\in C^1$ and $\inf_{v\ge 0}\gamma'(v)>-\infty$, then for $0<\tau_*<\inf_{x_0\in\r^d}\tau_{\min}(x_0)$ the periodic control~\eqref{eq.alg2+} provides~\eqref{eq.inf-u-sigma0} \emph{for any} initial condition.
In other words, the sampled-time emulation of the continuous feedback at a sufficiently high sampling rate ensures \emph{global} stability of the closed-loop system with a known convergence rate.
\end{remark}
\begin{remark}\label{rem.restrict}
The existing results on stability of nonlinear systems with sampled-time control~\eqref{eq.alg2+} typically adopt some continuity assumptions on the continuous-time controller. One of the standard assumptions~\cite{HsuSastry1987,Burlion2006} is the Lipschitz continuity of $\u(\cdot)$ and uniform boundedness of $F'_u(x,u)$. The weakest assumption of this type~\cite{Owens1990} requires\footnote{Notice that in~\cite{Owens1990}, the continuous-time system is exponentially stable with \emph{quadratic} Lyapunov function $V(x)$, whereas the sampled-time system is only asymptotically stable (without any explicit estimate for the convergence rate).} the map $(x,x_*)\mapsto F(x,\u(x_*))$ to be continuous (usually, $\u$ has to be continuous). 
\textcolor{black}{Theorem~\ref{thm.self} does not rely on any of these conditions, however, $|F(x,\u(x)|=O(|x|)$ as $|x|\to 0$ due to Assumptions~\ref{ass.gradient} and~\ref{ass.non-degen}. The latter condition fails to hold when the continuous-time control $u=\u(x)$ provides finite-time stabilization~\cite{BhatBernstein2000,MOULAY2006}. This agrees with Remark~\ref{rem.dwell-time}, explaining that our procedure of event-triggered controller design cannot guarantee local uniform dwell-time positivity in the latter case.} \end{remark}

The strong advantage of the self-triggered and the periodic sampling algorithms is the possibility to \emph{schedule} communication and control tasks. Such algorithms are more convenient for real-time embedded systems engineering than the event-triggered controller~\eqref{eq.alg1}, which requires constant monitoring of the solution $x(t)$ and potentially can use the communication channel at any time. The downside of this is the necessity to estimate the inter-sampling time $\tau(\cdot)$. The conservatism of such estimates leads to more data transmissions and control switchings than the event-triggered controller~\eqref{eq.alg1} needs.

\subsection{Periodic event-triggered stabilization}

A combination of the event-triggered and periodic sampling, inheriting the advantages of both approaches, is referred to as \emph{periodic} event-triggered control~\cite{HeemelsDonkers:2013,HeemelsDonkersTeel:2013}. Unlike usual event-triggered control, the triggering condition is checked \emph{periodically} with some fixed period $h>0$,
i.e. the control input can be recalculated only at time $kh$, where $k=0,1,\ldots$. This automatically excludes the possibility of Zeno behavior (obviously, $t_{n+1}-t_n\ge h>0$) and simplifies scheduling of the computational and communication tasks.

The main difficulty in designing the periodic event-triggered controller is to find such a triggering condition that its validity at time $kh$ automatically implies the desired control goal~\eqref{eq.inf-u-sigma0} on the interval $[kh,(k+1)h]$, even if the control input at time $t=kh$ remains unchanged.
Fixing two constants $\tilde\sigma\in (\sigma,1)$ and $K>1$, we introduce the boolean function (predicate)
\be\label{eq.trigger-periodic}
\begin{aligned}P(x,u)=
&W(x,u)<-\tilde\sigma\gamma(V(x))\\
&\land\quad\frac{|V'(x)|\,|F(x,u)|+|F(x,u)|^2}{M(x)|W(x,u)|}\leq K\\
\end{aligned}
\ee
Here $M(x)$ is the function from~\eqref{eq.non-degen1}. The conditions~\eqref{eq.inf-u-gamma} and~\eqref{eq.non-degen1} imply that $P(x_*,\u(x_*))$ is true for any $x_*\ne 0$ since
\be\label{eq.auxaux}
W(x_*,\u(x_*))\le -\gamma(V(x_*))<-\tilde\sigma\gamma(V(x_*)).
\ee

Choosing the sampling period $h>0$ in a way specified later (Lemma~\ref{lem.key-lemma1}), the following key property can be guaranteed: if $P(x(t_*),u_*)$ holds for some $t_*$ then the static control input $u(t)\equiv u_*$ provides the validity of~\eqref{eq.inf-u-sigma0} for $t\in [t_*,t_*+h)$ (notice that $P(x(t),u_*)$ need not be true on this interval). This suggests the following periodic event-triggered algorithm. At the initial instant $t_0=0$, calculate the control input $u_0\dfb\u(x(t_0))$. If $x(t_0)=0$, we may freeze the control input $u(t)\equiv u_0\quad\forall t\ge 0$. At any time $t=kh$, where $k=1,2,\ldots,$, the condition $P(x(kh),u_0)$ is checked, until one finds the first $k_1\ge 1$ such that $P(x(k_1h),u_0)$ is false. At the instant $t_1=k_1h$, the control input is switched to $u_1=\u(x(t_1))$, and the procedure is repeated again: if $x(t_1)=0$, one can freeze $u(t)\equiv u_1$, otherwise, $u(t)=u_1$ until the first instant $k_2h$ (with $k_2>k_1$), where $P(x(k_2h),u_1)$ is false, and so on. Mathematically, the algorithm is as follows
\be\label{eq.alg3}
\begin{gathered}
u(t)=u_n\dfb\u(x(k_nh))\;\;\forall t\in [k_nh,k_{n+1}h);\quad k_0=0,\\
k_{n+1}=\begin{cases}\min\left\{k>k_n: \lnot P(x(kh),u_n) \right\},\,&x(k_nh)\ne 0,\\
\infty,\,&x(k_nh)=0.
\end{cases}
\end{gathered}
\ee
(by definition, $\min\emptyset=+\infty$).

Notice that the algorithm~\eqref{eq.alg3} implicitly depends on three parameters: $\sigma\in (0,1)$, $\tilde\sigma\in(\sigma,1)$ and $K>1$. The role of the first parameter is the same as in Algorithm~\eqref{eq.alg1} (it regulates the converges rate). The parameters $\tilde\sigma$ and $K$ determine the maximal sampling period $h$: the less restrictive condition $P(x(kh),u_n)$ is, the more often it has to be checked in order to guarantee the desired inter-sampling behavior, as will be explained in more detail in Remark~\ref{rem.sigma-K-remark}.

The choice of $h>0$ is based on the following lemma, similar to Lemma~\ref{lem.key-lemma} and dealing
with the solution $\xi(t)=\xi(t|\bar x,u_*)$ to the Cauchy problem~\eqref{eq.cauchy}. Unlike Lemma~\ref{lem.key-lemma}, $u_*\ne\u(\bar x)$.
\begin{lemma}\label{lem.key-lemma1}
Let Assumptions~\ref{ass.F}-\ref{ass.non-degen} be valid, $\gamma(\cdot)$ be either non-decreasing or $C^1$-smooth,
$\tilde\sigma\in (\sigma,1)$ and $K>1$. Then there exists a function $\tau^0:\r^d\to (0,\infty)$ such that
\begin{enumerate}
\item $\tau^0$ is uniformly positive on any compact set;
\item if $x_*\ne 0$, $\bar x\in B(x_*)$ and $P(\bar x,\u(x_*))$ is valid, then the solution $\xi(t)=\xi(t|\bar x,\u(x_*))$ is well-defined for $t\in [0,\tau^0(x_*)]$ and
the following inequality holds
\be\label{eq.w-ineq-main+}
W(\xi(t),\u(x_*))<-\sigma \gamma(V(\xi(t)))\quad\forall t<\tau^0(x_*).
\ee
\end{enumerate}
If the functions $\vk,\nu,M$ are \emph{globally} bounded, $\gamma\in C^1$ and $\inf\limits_{v\ge 0}\gamma'(v)>-\infty$, then $\tau^0$ is globally uniformly positive.
\end{lemma}

Lemma~\ref{lem.key-lemma1} is proved in Appendix~\ref{app.proof}, where an explicit formula for $\tau^0(\cdot)$ is found. This lemma entails the following result.
\begin{theorem}\label{thm.periodic}
Let the assumptions of Lemma~\ref{lem.key-lemma1} be valid. For any compact set $\mathcal K\subset\r^d$, choose the sampling interval
$h\in\left(0,\inf\limits_{x\in B(\mathcal K)}\tau^0(x)\right)$. Then the periodic event-triggered controller~\eqref{eq.alg3} provides the inequality~\eqref{eq.inf-u-sigma0} for any $x(0)\in\mathcal {K}$. If the functions $\vk,\nu,M$ are globally bounded, $\gamma\in C^1$ and $\inf\limits_{v\ge 0}\gamma'(v)>-\infty$, then the controller~\eqref{eq.alg3} provides~\eqref{eq.inf-u-sigma0} for \emph{any} $x(0)\in\r^d$ whenever $h<\inf_{\r^d}\tau^0$.
\end{theorem}
\begin{proof}
Via induction on $k=0,1,\ldots,$, we are going to prove that~\eqref{eq.inf-u-sigma0} holds on $[kh,(k+1)h)$ (in particular, the solution remains bounded between two sampling instants). The induction base $k=0$ is immediate from Lemma~\ref{lem.key-lemma1} and the definition of $h$. Since $h\le\tau^0(x(0))$ and $P(x(0),\u(x(0)))$ holds thanks to~\eqref{eq.auxaux}, the solution $x(t)=\xi(t|x_0,u_0)$ satisfies~\eqref{eq.inf-u-sigma0} due to~\eqref{eq.w-ineq-main+}. To prove the induction step, suppose that the statement has been proved
for $k\le\bar k-1$, in particular, $V(x(t))$ is non-increasing for $t\in[0,\bar kh)$. By construction of the algorithm, the condition $P(x(kh),u(kh))$ is true, where $u(kh)=\u(x(k_nh))$ and $k_n\le k$ (no matter if the control is recalculated at $t=kh$ or not). Applying Lemma~\ref{lem.key-lemma1} to $x_*=x(k_nh)$ and $\bar x=x(\bar kh)\in B(x_*)$, one obtains that the solution $x(t)=\xi(t-kh|\bar x,\u(x_*))$ satisfies~\eqref{eq.inf-u-sigma0} for $t\in[\bar kh,(\bar k+1)h)$ since $x_*\in B(x(0))$ and therefore $h\le\tau^0(x_*)$. This proves the induction step.
\end{proof}
\begin{remark}\label{rem.sigma-K-remark}
Obviously, the condition $P(x,u)$ is the less restrictive, the smaller is $(\tilde\sigma-\sigma)$ and the greater is $K>1$. It can be seen, however (see Appendix~\ref{app.proof}) that when
$\tilde\sigma\to\sigma$ or $K\to\infty$, one has $\tau^0(x_*)\to 0$, i.e. the periodic event-triggered algorithm reduces to the usual event-triggered algorithm~\eqref{eq.alg1}, continuously monitoring the state. The case where $K\to 1$ and $\tilde\sigma\to 1$ corresponds to the most restrictive
condition $P(x,u)$. In this case, as can be shown, $\tau^0(x_*)\to \tau(x_*)$ from Lemma~\ref{lem.key-lemma}, and hence $\tau(\mathcal K)\to\min\limits_{x_0\in\mathcal K}\tau_{\min}(x_0)$.
In the worst-case choice of $x_0\in\mathcal K$, the algorithm~\eqref{eq.alg3} behaves as the special case of time-triggered control~\eqref{eq.alg2+} with $t_{n+1}-t_n=\tau_{\min}(x_0)$.
\end{remark}

\section{Numerical Examples}\label{sec.example}

In this section, two examples illustrating the applications of algorithm~\eqref{eq.alg1} are considered.

\subsection{Event-triggered backstepping for cruise control}

\begin{figure}[h]
\center
\includegraphics[width=\columnwidth]{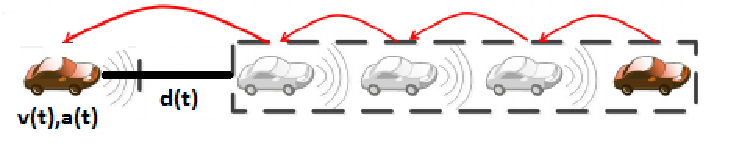}
\caption{Merging to a platoon}\label{fig.2}
\end{figure}

Our first example illustrates the procedure of event-triggered backstepping with guaranteed dwell-time positivity in the following problem, regarding the design of full-range, or stop and go, adaptive cruise control (ACC) systems~\cite{AmesTabuada:2017,Babu:2016,WangRiender:2016}.
The main purpose of ACC systems is to adjust automatically the vehicle speed to maintain a safe distance from vehicles ahead (the distance to the predecessor vehicle, as well as its velocity, is measured by onboard radars, laser sensors or cameras). We consider, however, a more general problem that can be solved by ACC, namely, keeping the predefined distance to the predecessor vehicle. Such a problem is natural e.g. when the vehicle has to safely merge a \emph{platoon} of vehicles (Fig.~\ref{fig.2}), move in a platoon or leave it~\cite{Chien:1995}. In the simplest situation the platoon travels at constant speed $v_0>0$. Denoting and the the desired distance from the vehicle to the platoon by $d_0$, the control goal is formulated as follows
\be\label{eq.merge}
d(t)-d_0\xrightarrow[t\to\infty]{} 0,\quad v(t)-v_0\xrightarrow[t\to\infty]{} 0.
\ee

We consider the standard third-order model of a vehicle's longitudinal dynamics~\cite{DolkPloegHeemels:2017,DolkBorgersHeemels:2017}
\be\label{eq.vehi}
\tau(v)\dot a(t)+a(t)=u(t),\quad a(t)=\dot v(t).
\ee
Here $a(t)$ is the controller vehicle's actual acceleration, whereas $u(t)$ can be treated as the commanded (desired) acceleration. The function $\tau(v)$ depends on the dynamics of the servo-loop and characterizes the driveline constant, or time lag between the commanded and actual accelerations. We suppose the function $\tau(v)$ to be known, the vehicle being able to measure $d(t),v(t),a(t)$ and aware of the platoon's speed $v_0$.

To design an exponentially stabilizing CLF in this problem, we use the well-known backstepping procedure~\cite{KrsticKokotovicBook,Khalil}. We introduce the functions $x_1,x_2,x_3$ as follows
\ben
\begin{aligned}
x_1(t)&\dfb d(t)-d_0\Longrightarrow \\
x_2(t)&\dfb \dot x_1(t)+kx_1(t)=(v_0-v(t))+kx_1(t)\\
x_3(t)&\dfb \dot x_2(t)+kx_2(t)= -a(t)+2k(v_0-v(t))+k^2x_1(t).
\end{aligned}
\een

\begin{figure*}[ht]
\centering
\begin{subfigure}[b]{0.24\textwidth}
\center
\includegraphics[width=\columnwidth]{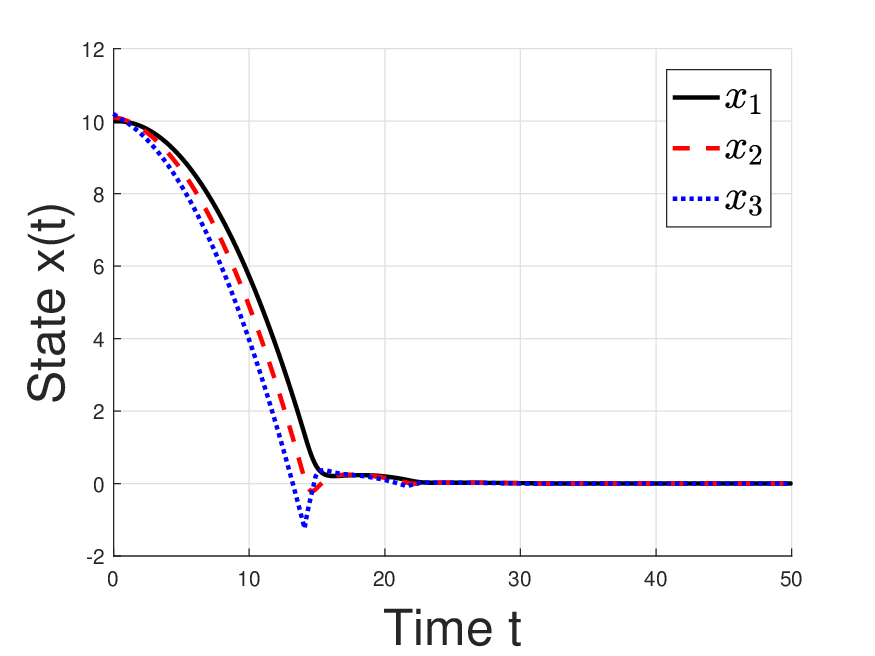}
\end{subfigure}
\begin{subfigure}[b]{0.24\textwidth}
\center
\includegraphics[width=\columnwidth]{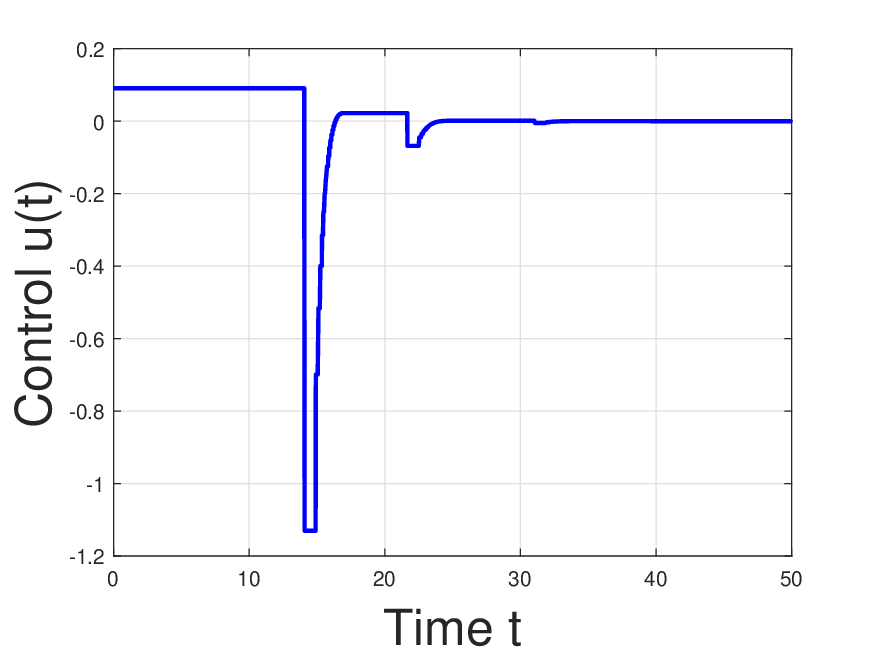}
\end{subfigure}
\begin{subfigure}[b]{0.24\textwidth}
\center
\includegraphics[width=\columnwidth]{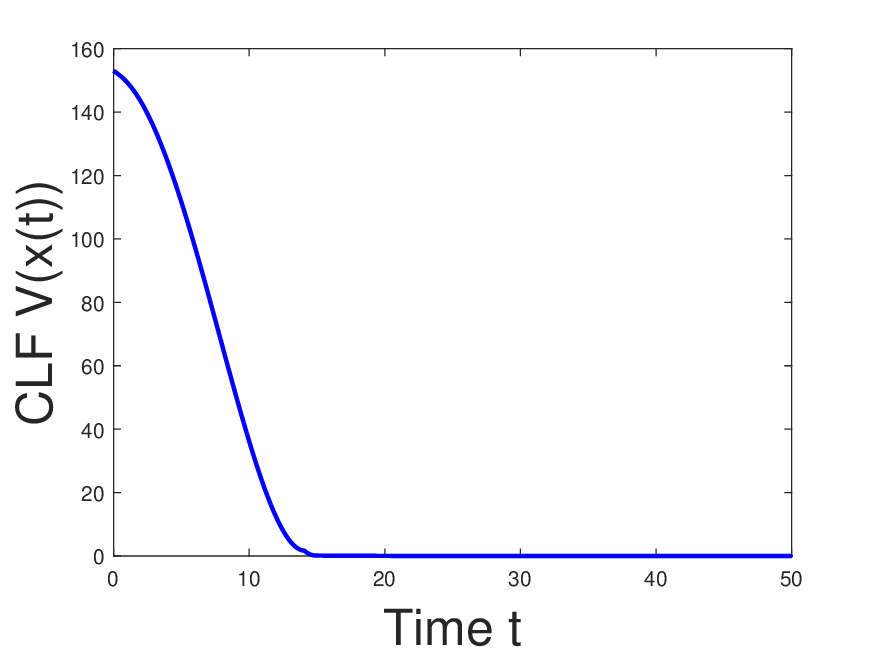}
\end{subfigure}
\begin{subfigure}[b]{0.24\textwidth}
\center
\includegraphics[width=\columnwidth]{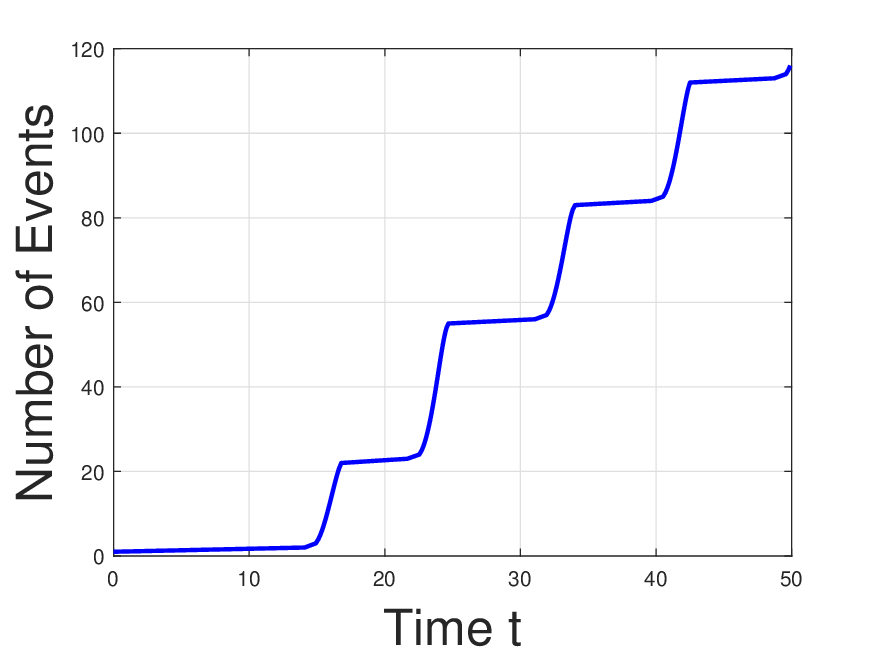}
\end{subfigure}
\begin{subfigure}[b]{0.24\textwidth}
\center
\includegraphics[width=\columnwidth]{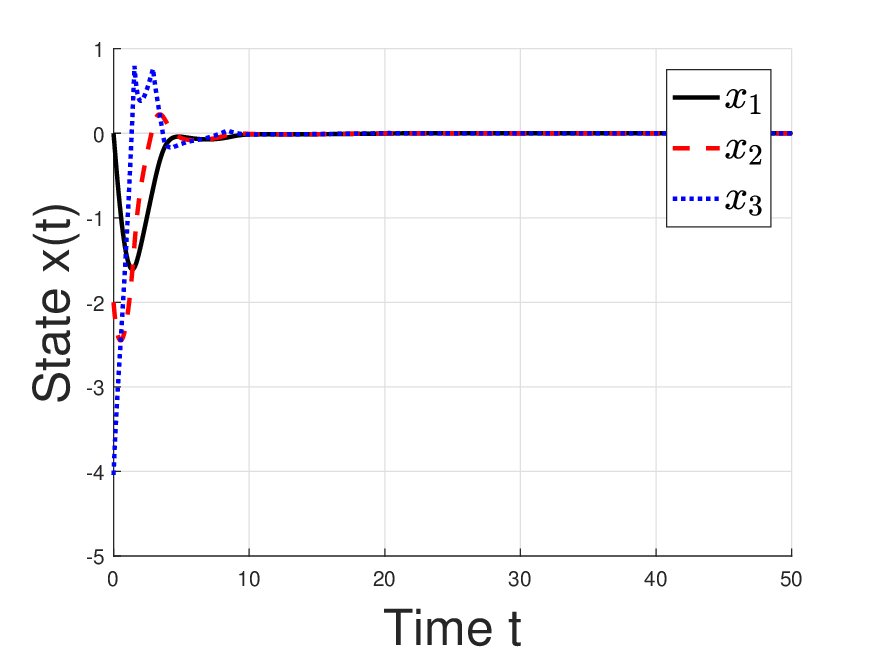}
\end{subfigure}
\begin{subfigure}[b]{0.24\textwidth}
\center
\includegraphics[width=\columnwidth]{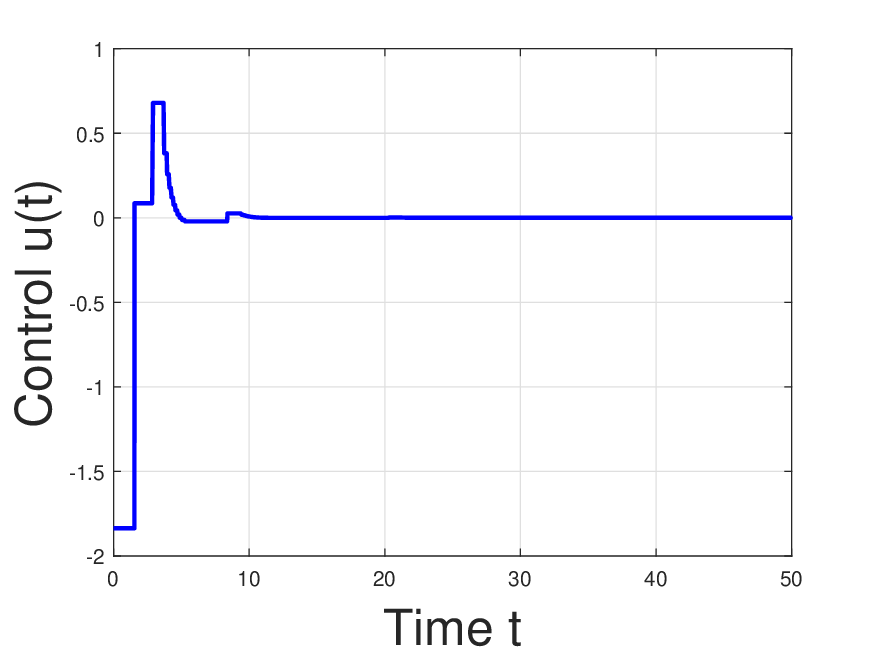}
\end{subfigure}
\begin{subfigure}[b]{0.24\textwidth}
\center
\includegraphics[width=\columnwidth]{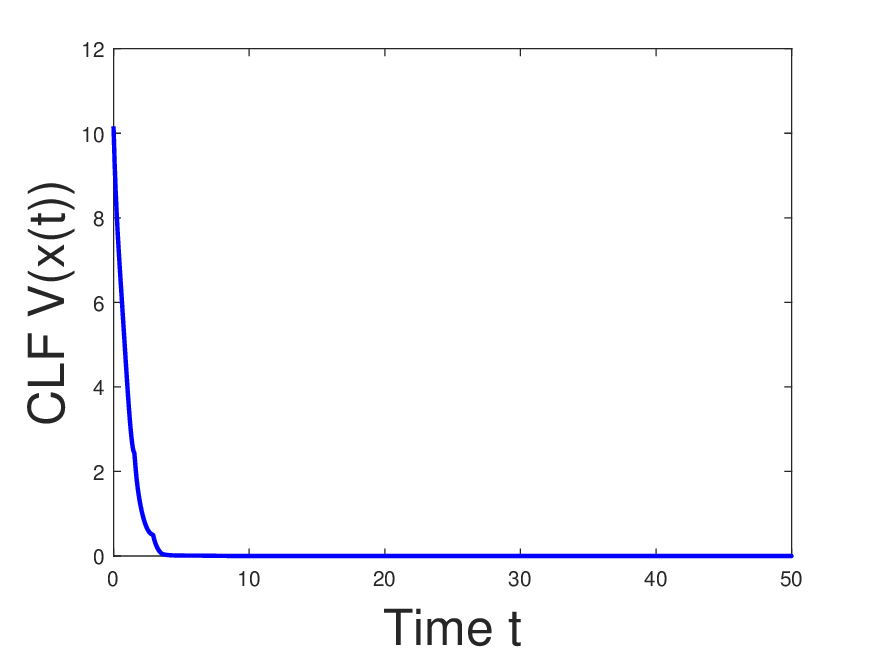}
\end{subfigure}
\begin{subfigure}[b]{0.24\textwidth}
\center
\includegraphics[width=\columnwidth]{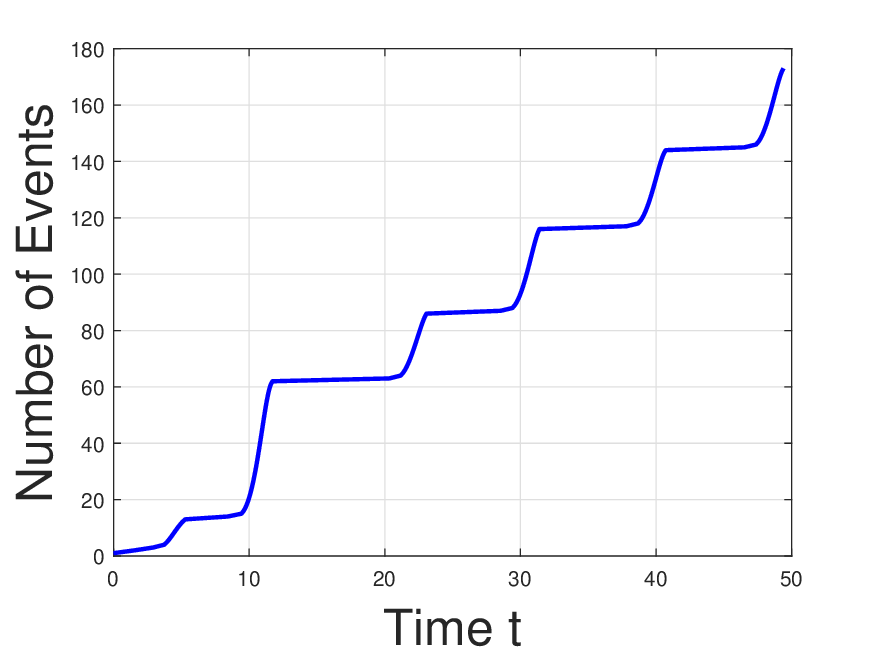}
\end{subfigure}
\caption{Event-triggered stabilization of system~\eqref{eq.vehi1}}\label{Fig.ex1}
\end{figure*}
By noticing that $v_0-v(t)=x_2-kx_1$ and $a(t)=2kx_2(t)-k^2x_1(t)-\xi_3(t)$, the equations~\eqref{eq.vehi} are rewritten as follows
\be\label{eq.vehi1}
\begin{aligned}
\dot x_1&=x_2-kx_1\\
\dot x_2&=x_3-kx_2\\
\dot x_3&=k^2[x_2-kx_1]+\\&+[\tau(v)^{-1}-2k](2kx_2-k^2x_1-x_3)-\tau(v)^{-1}u\\
v&=v_0-(x_2-kx_1).
\end{aligned}
\ee
It can be easily shown now that $V(x)=\frac{1}{2}(x_1^2+x_2^2+x_3^2)$ is the CLF for the system~\eqref{eq.vehi1}
whenever $k>1$, corresponding to the feedback controller $\u(x)$ as follows
\[
\begin{aligned}
\u(x)&\dfb\tau(v)k^2[x_2-kx_1]+\\&+[1-2k\tau(v)](2kx_2-k^2x_1-x_3)-\tau(v)(x_1-kx_3).
\end{aligned}
\]
Indeed, a straightforward computation shows that
\[
\begin{aligned}
F(x,\u(x))&=(x_2-kx_1,x_3-kx_2,x_1-kx_3)^{\top},\\
V'(x)F(x,\u(x))&=-2(k-1)V(x)-\\&-\frac{1}{2}[(x_1-x_2)^2+(x_1-x_3)^2+(x_2-x_3)^2],
\end{aligned}
\]
entailing~\eqref{eq.es-clf} with $\ae=2(k-1)$. It can be easily shown that all assumptions of Theorem~\ref{thm.dwell} hold. The algorithm~\eqref{eq.alg1} gives an event-triggered ACC algorithm.

In Fig.~\ref{Fig.ex1}, we simulate the behavior of the algorithm~\eqref{eq.alg1} with $\sigma=0.9$, choosing $k=1.01$ and $\tau=0.3s$ for two situations. In the first situation (plots on top)
the vehicle initially travels with the same speed as the platoon ($v(0)-v_0=0$), but needs to decrease the distance by $10$m, i.e. $x_1(0)=d(0)-d_0=10$, $x_2(0)=kx_1(0)$, $x_3(0)=k^2x_1(0)$. In the second case (plots at the bottom), the vehicle needs to decrease its speed by $2$m/s, keeping the initial distance to the platoon: $d_0=d(0)$, $v(0)-v_0=2$,
and thus $x_1(0)=0,x_2(0)=v_0-v(0)=-2,x_3(0)=-4k$. One may notice that the vehicle's trajectories periods of ``harsh'' braking, which cause discomfort of the human occupants (as well as large values of the jerk, caused by rapid switch of the control input). In this simple example, intended for demonstration of the design procedure, we do not consider these constraints.

\textcolor{black}{One may notice that in both situations the algorithm produces ``packs'' of 15-30 close events. In the first case, events are fired starting from $t_1=14.1$s, the maximal time elapsed between consecutive events is $6.38$s and the minimal time is $0.05$s. The average frequency of events is 3.2Hz. In the second case, the first event occurs at $t_1=1.5s$, the maximal time between events is $8.6$s, the minimal time is $0.04$s. The average frequency of events is $3.6$Hz.}

\subsection{An example of non-exponential stabilization}

\begin{figure*}[ht]
\centering
\begin{subfigure}[t]{0.32\textwidth}
\center
\includegraphics[width=\columnwidth]{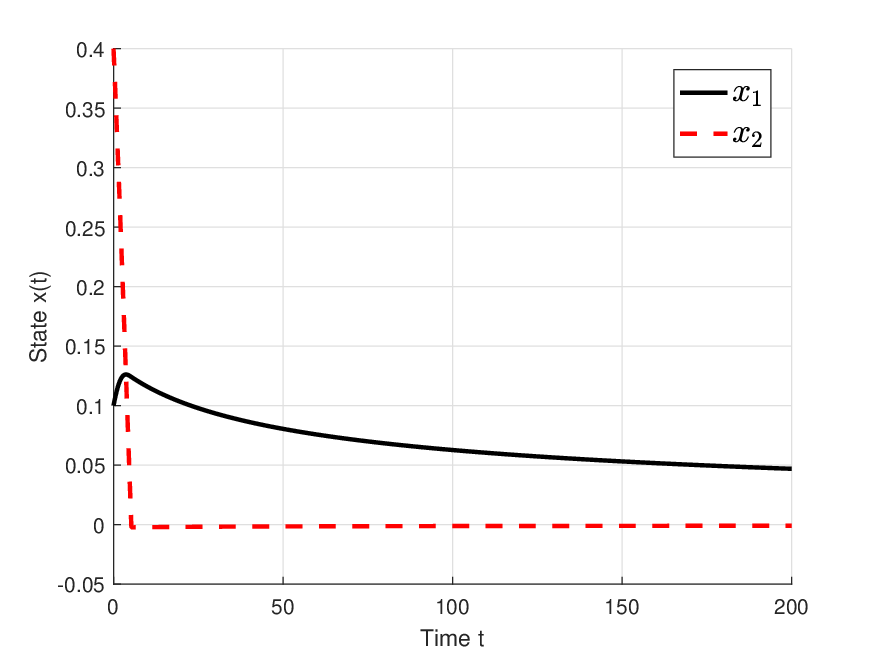}
\end{subfigure}
\begin{subfigure}[t]{0.32\textwidth}
\center
\includegraphics[width=\columnwidth]{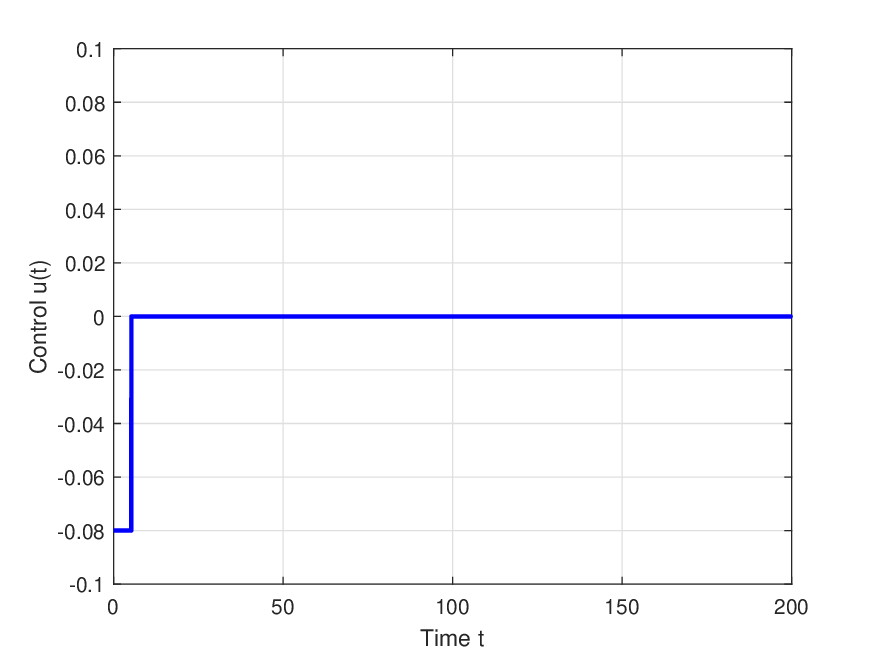}
\end{subfigure}
\begin{subfigure}[t]{0.32\textwidth}
\center
\includegraphics[width=\columnwidth]{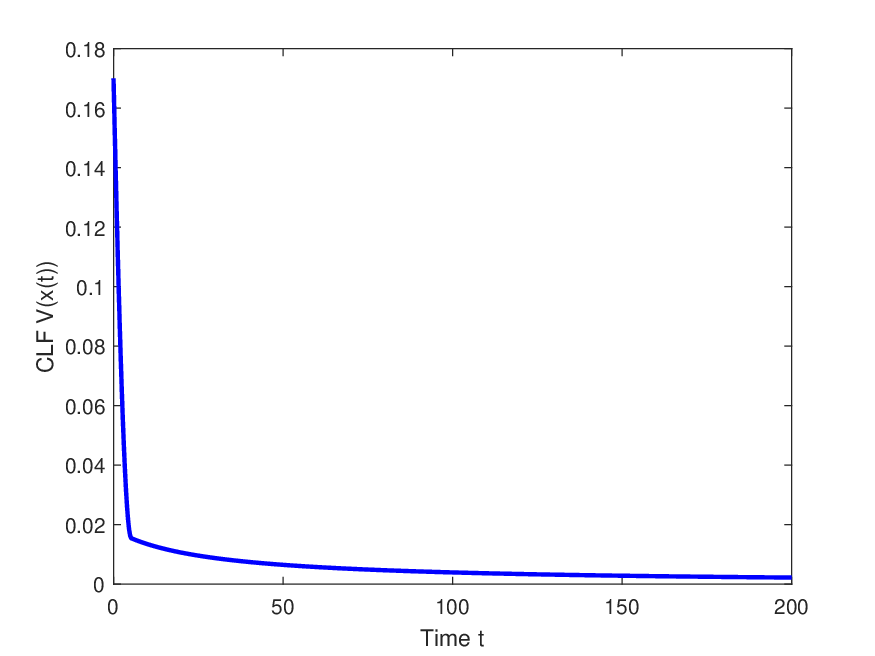}
\end{subfigure}
\caption{Event-triggered stabilization of system~\eqref{eq.syst-anta}}\label{Fig.ex2}
\end{figure*}
Our second example is borrowed from~\cite{AntaTabuada:2008} and deals with a two-dimensional homogeneous system
\be\label{eq.syst-anta}
\begin{gathered}
\dot x_1=-x_1^3+x_1x_2^2,\\
\dot x_2=x_1x_2^2+u-x_1^2x_2
\end{gathered}
\ee

The quadratic form $V(x)=\frac{1}{2}[x_1^2+x_2^2]$ satisfies~\eqref{eq.inf-u-gamma} with $\gamma(v)=v^2$ and $\u(x)=-x_2^3-x_1x_2^2$ since
\[
V'(x)F(x,\u(x))=-x_1^4-x_2^4\le -V^2/2.
\]
Therefore, the event-triggered algorithm~\eqref{eq.alg1} provides stabilization with convergence rate
\[
V(x(t))\le \left[V(x(0))+\sigma t/2\right]^{-1}.
\]

To compare our algorithm with the one reported in~\cite{Marchand:2013} and based on the Sontag controller, we simulate the behavior of the system for $x_1(0)=0.1, x_2(0)=0.4$, choosing $\sigma=0.9$. The results of numerical simulation (Fig.~\ref{Fig.ex2}) are similar to those presented in~\cite{Marchand:2013}. Although the convergence of the solution is slow ($V(x(t))=O(t^{-1})$ and $|x(t)|=O(t^{-1/2})$), its second component and  the control input converge very fast. 
\textcolor{black}{During the first $200$s, only two events are detected at times $t_0=0$ and $t_1\approx 5.26$, after which the control is fixed at $u(t)\approx -6\cdot 10^{-7}$.}

\section{Conclusion}\label{sec.concl}

In this paper, we address the following fundamental question: let a nonlinear system admit a control Lyapunov function (CLF), corresponding to a continuous-time stabilizing controller with a certain (e.g. exponential or polynomial) convergence rate. Does this imply the existence of an event-triggered controller, providing the same convergence rate? Under certain natural assumptions, we give an affirmative answer and show that such a controller in fact also provides the positive dwell time between consecutive events. Moreover, we show that if the initial condition is confined to a known compact set, this problem can be also solved by self-triggered and periodic event-triggered controllers.
Our results can also be extended to robust control Lyapunov functions (RCLF), extending the concept of CLF to systems with disturbances.

Analysis of the proofs reveals that the main results of the paper retain their validity in the case where the CLF is proper yet not positive definite, and its \emph{compact} zero set $X_0=\{x\in\r^d: V(x)=0\}$ consists of the equilibria of the system~\eqref{eq.closed-loop}. If our standing assumptions hold, then algorithms~\eqref{eq.alg1},\eqref{eq.alg2},\eqref{eq.alg2+},\eqref{eq.alg3} provide that $V(x(t))\xrightarrow[t\to\infty]{}0$ (with a known convergence rate) and any solution converges to $X_0$ in the sense that ${\rm dist}(x(t),X_0)\xrightarrow[t\to\infty]{} 0$.
At the same time, Lyapunov stabilization of unbounded sets (e.g. hyperplanes~\cite{Ames:2014}) requires additional assumptions on CLFs; the relevant extensions are beyond the scope of this paper.
\color{black}

Although the existence of CLFs can be derived from the inverse Lyapunov theorems, to find a CLF satisfying Assumptions~\ref{ass.F}-\ref{ass.non-degen} can in general be non-trivial; computational approaches to cope with it are subject of ongoing research. Especially challenging are problems of \emph{safety-critical} control, requiring to design a control Lyapunov-barier function (CLBF). Other important problems are event-triggered and self-triggered redesign of dynamic continuous-time controllers (needed e.g. when the state vector cannot be fully measured) and stabilization with non-smooth CLFs~\cite{McConley:1998}. 

\bibliographystyle{IEEETRan}
\bibliography{event,nonlinear}

\begin{thebibliography}{10}
\providecommand{\url}[1]{#1}
\csname url@samestyle\endcsname
\providecommand{\newblock}{\relax}
\providecommand{\bibinfo}[2]{#2}
\providecommand{\BIBentrySTDinterwordspacing}{\spaceskip=0pt\relax}
\providecommand{\BIBentryALTinterwordstretchfactor}{4}
\providecommand{\BIBentryALTinterwordspacing}{\spaceskip=\fontdimen2\font plus
\BIBentryALTinterwordstretchfactor\fontdimen3\font minus
  \fontdimen4\font\relax}
\providecommand{\BIBforeignlanguage}[2]{{%
\expandafter\ifx\csname l@#1\endcsname\relax
\typeout{** WARNING: IEEEtran.bst: No hyphenation pattern has been}%
\typeout{** loaded for the language `#1'. Using the pattern for}%
\typeout{** the default language instead.}%
\else
\language=\csname l@#1\endcsname
\fi
#2}}
\providecommand{\BIBdecl}{\relax}
\BIBdecl

\bibitem{ProMazo_HSCC}
A.~Proskurnikov and M.~Mazo~Jr., ``Lyapunov design for event-triggered
  exponential stabilization,'' in \emph{HSCC'18: 21st International Conference
  on Hybrid Systems: Computation and Control (part of CPSWeek)}, 2018.

\bibitem{KalmanBertram:1960}
R.~Kalman and J.~Bertram, ``Control system analysis and design via the
  ``{S}econd {M}ethod'' of {L}yapunov. {I.} {C}ontinuous-time systems,''
  \emph{Journal of Basic Engineering}, vol.~32, pp. 371--393, 1960.

\bibitem{Artstein:1983}
Z.~Artstein, ``Stabilization with relaxed control,'' \emph{Nonlinear Analysis.
  Theory, Methods \& Applications}, vol.~7, no.~11, pp. 1163--1173, 1983.

\bibitem{Sontag:1989}
E.~Sontag, ``A ``universal'' construction of {A}rtstein's theorem on nonlinear
  stabilization,'' \emph{Syst. Control Lett.}, vol.~13, pp. 117--123, 1989.

\bibitem{Rantzer:2001}
A.~Rantzer, ``A dual to {L}yapunov's stability theorem,'' \emph{Syst. Control
  Lett.}, vol.~42, pp. 161--168, 2001.

\bibitem{FauborgPomet:2000}
L.~Faubourg and J.-B. Pomet, ``Control {L}yapunov functions for homogeneous
  {J}urdjevic-{Q}uinn systems,'' \emph{ESAIM: Control, Optim., Calculus
  Variations}, vol.~5, pp. 293--311, 2000.

\bibitem{KokotovicArcak:2001}
P.~V. Kokotovi{\'c} and M.~Arcak, ``Constructive nonlinear control: A
  historical perspective,'' \emph{Automatica}, vol.~37, no.~5, pp. 637--662,
  2001.

\bibitem{Khalil}
H.~Khalil, \emph{Nonlinear systems}.\hskip 1em plus 0.5em minus 0.4em\relax
  Englewood Cliffs, NJ: Prentice-Hall, 1996.

\bibitem{Praly:1991}
L.~Praly, B.~{d'A}ndr{\'e}a Novel, and J.-M. Coron, ``Lyapunov design of
  stabilizing controllers for cascaded systems,'' \emph{IEEE Trans. Autom.
  Control}, vol.~36, no.~10, pp. 1177--1181, 1991.

\bibitem{KrsticKokotovicBook}
M.~Krsti{\'c}, I.~Kanellakopoulos, and P.~Kokotovi{\'c}, \emph{Nonlinear and
  adaptive control design}.\hskip 1em plus 0.5em minus 0.4em\relax Wiley, 1995.

\bibitem{SepulchreKokotovicBook}
R.~Sepulchre, M.~Jankovi{\'c}, and P.~Kokotovi{\'c}, \emph{Robust Nonlinear
  Control Design. State-Space and Lyapunov Techniques}.\hskip 1em plus 0.5em
  minus 0.4em\relax Springer London, 1997.

\bibitem{WangLiu:2017}
Z.~Wang, X.~Liu, K.~Liu, S.~Li, and H.~Wang, ``Backstepping-based {L}yapunov
  function construction using approximate dynamic programming and sum of square
  techniques,'' \emph{IEEE Trans. Cybern.}, vol.~47, no.~10, pp. 3393--3403,
  2017.

\bibitem{Furqon:2017}
R.~Furqon, Y.-J. Chen, M.~Tanaka, K.~Tanaka, and H.~Wang, ``An {SOS}-based
  control lyapunov function design for polynomial fuzzy control of nonlinear
  systems,'' \emph{IEEE Trans. Fuzzy Syst.}, vol.~25, no.~4, pp. 775--787,
  2017.

\bibitem{VerdierMazo:2017}
C.~Verdier and M.~Mazo~Jr, ``Formal controller synthesis via genetic
  programming,'' \emph{IFAC-PapersOnLine}, vol.~50, no.~1, pp. 7205--7210,
  2017.

\bibitem{FreemanKokotovic:1996}
R.~Freeman and P.~Kokotovi{\'c}, ``Inverse optimality in robust
  stabilization,'' \emph{{SIAM} J. Control Optim.}, vol.~34, pp. 1365--1391,
  1996.

\bibitem{Camilli:2008}
F.~Camilli, L.~Gr\"une, and F.~Wirth, ``Control {L}yapunov functions and
  {Z}ubov's method,'' \emph{SIAM J. Control Optim.}, vol.~47, no.~1, pp.
  301--326, 2008.

\bibitem{FreemanKokotovicBook}
R.~Freeman and P.~Kokotovi{\'c}, \emph{Robust Nonlinear Control Design.
  State-Space and Lyapunov Techniques}.\hskip 1em plus 0.5em minus 0.4em\relax
  Birkh\"auser, 1996.

\bibitem{KellettTeel:2004}
C.~Kellett and A.~Teel, ``Discrete-time asymptotic controllability implies
  smooth control {L}yapunov function,'' \emph{Syst. Control Lett.}, vol.~52,
  pp. 349--59, 2004.

\bibitem{Jancovic:2001}
M.~Jancovi{\'c}, ``Control {L}yapunov-{R}azumikhin functions and robust
  stabilization of time delay systems,'' \emph{IEEE Trans. Autom. Control},
  vol.~46, no.~7, pp. 1048--1060, 2001.

\bibitem{Sanfelice:2013}
R.~Sanfelice, ``On the existence of control {L}yapunov functions and
  state-feedback laws for hybrid systems,'' \emph{IEEE Trans. Autom. Control},
  vol.~58, no.~12, pp. 3242--3248, 2013.

\bibitem{Ames:2014}
A.~Ames, K.~Galloway, K.~Sreenath, and J.~Grizzle, ``Rapidly exponentially
  stabilizing control {L}yapunov functions and hybrid zero dynamics,''
  \emph{IEEE Trans. Autom. Control}, vol.~59, no.~4, pp. 876--891, 2014.

\bibitem{RomdlonyJayawardhana:2016}
M.~Romdlony and B.~Jayawardhana, ``Stabilization with guaranteed safety using
  control {L}yapunov--barrier function,'' \emph{Automatica}, vol.~66, pp.
  39--47, 2016.

\bibitem{NilssonTabuada:2016}
P.~Nilsson, O.~Hussien, A.~Balkan, Y.~Chen, A.~Ames, J.~Grizzle, N.~Ozay,
  H.~Peng, and P.~Tabuada, ``Correct-by-construction adaptive cruise control:
  Two approaches,'' \emph{IEEE Trans. Control Syst. Tech.}, vol.~24, no.~4, pp.
  1294--1307, 2016.

\bibitem{AmesTabuada:2017}
A.~Ames, X.~Xu, J.~Grizzle, and P.~Tabuada, ``Control barrier function based
  quadratic programs for safety critical systems,'' \emph{IEEE Trans. Autom.
  Control}, vol.~62, no.~8, pp. 3861--3876, 2017.

\bibitem{Hetel2017}
L.~Hetel, C.~Fiter, H.~Omran, A.~Seuret, E.~Fridman, J.-P. Richard, and S.~I.
  Niculescu, ``Recent developments on the stability of systems with aperiodic
  sampling: An overview,'' \emph{Automatica}, vol.~76, pp. 309 -- 335, 2017.

\bibitem{NesicTeelKokotovich:1999}
D.~Ne{\v s}i{\'c}, A.~Teel, and P.~Kokotovi{\' c}, ``Sufficient conditions for
  stabilization of sampled-data nonlinear systems via discrete-time
  approximations,'' \emph{Syst. Control Lett.}, vol.~38, no. 4-5, pp. 259--270,
  1999.

\bibitem{NesicTeel:2004}
D.~Ne{\v s}i{\'c} and A.~Teel, ``A framework for stabilization of nonlinear
  sampled-data systems based on their approximate discrete-time models,''
  \emph{IEEE Trans. Autom. Control}, vol.~49, no.~7, pp. 1103--1122, 2004.

\bibitem{ArcakNesic:2004}
M.~Arcak and D.~Ne{\v s}i{\'c}, ``A framework for nonlinear sampled-data
  observer design via approximate discrete-time models and emulation,''
  \emph{Automatica}, vol.~40, no.~11, pp. 1931--1938, 2004.

\bibitem{Astrom:2002}
K.~Astr\"om and B.~Bernhardsson, ``Comparison of {R}iemann and {L}ebesgue
  sampling for first order stochastic systems,'' in \emph{Proc. of IEEE Conf.
  Decision and Control}, Las Vegas, 2002, pp. 2011--2016.

\bibitem{Tabuada:2007}
P.~Tabuada, ``Event-triggered real-time scheduling of stabilizing control
  tasks,'' \emph{IEEE Trans. Autom. Control}, vol.~52, no.~9, pp. 1680--1685,
  2007.

\bibitem{BorgersHeemels:2014}
D.~Borgers and W.~Heemels, ``Event-separation properties of event-triggered
  control systems,'' \emph{IEEE Trans. Autom. Control}, vol.~59, no.~10, pp.
  2644--2656, 2014.

\bibitem{Araujo:2014}
J.~Araujo, M.~Mazo, A.~Anta, P.~Tabuada, and K.~Johansson, ``System
  architectures, protocols and algorithms for aperiodic wireless control
  systems,'' \emph{IEEE Trans. Ind. Inform.}, vol.~10, no.~1, pp. 175--184,
  2014.

\bibitem{Postoyan2015}
R.~Postoyan, P.~Tabuada, D.~Ne{\v s}i{\'c}, and A.~Anta, ``A framework for the
  event-triggered stabilization of nonlinear systems,'' \emph{IEEE Transactions
  on Automatic Control}, vol.~60, no.~4, pp. 982--996, 2015.

\bibitem{Goebel:2009}
R.~Goebel, R.~Sanfelice, and A.~Teel, ``Hybrid dynamical systems,'' \emph{IEEE
  Contr. Syst. Mag.}, vol.~29, no.~2, pp. 28--93, 2009.

\bibitem{DolkBorgersHeemels:2017}
V.~Dolk, D.~Borgers, and W.~Heemels, ``Output-based and decentralized dynamic
  event-triggered control with guaranteed $l_p$-gain performance and
  {Z}eno-freeness,'' \emph{IEEE Trans. Autom. Control}, vol.~62, no.~1, pp.
  34--39, 2017.

\bibitem{SelivanovFridman:2016}
A.~Selivanov and E.~Fridman, ``Event-triggered {$H_{\infty}$}-control: A
  switching approach,'' \emph{IEEE Trans. Autom. Control}, vol.~61, no.~10, pp.
  3221--3226, 2016.

\bibitem{YueTianHan:2013}
D.~Yue, E.~Tian, and Q.-L. Han, ``A delay system method for designing
  event-triggered controllers of networked control systems,'' \emph{IEEE Trans.
  Autom. Control}, vol.~58, no.~2, pp. 475--481, 2013.

\bibitem{SelivanovFridman:2016-1}
A.~Selivanov and E.~Fridman, ``Distributed event-triggered control of diffusion
  semilinear {PDEs},'' \emph{Automatica}, vol.~68, pp. 344--351, 2016.

\bibitem{LiuLiuDou:2014}
B.~Liu, D.-N. Liu, and C.-X. Dou, ``Exponential stability via event-triggered
  impulsive control for continuous-time dynamical systems,'' in \emph{Proc.
  Chinese Control Conf.}, 2014, pp. 4346--4350.

\bibitem{SeuretPrieurMarchand:2013}
A.~Seuret, C.~Prieur, and N.~Marchand, ``Stability of nonlinear systems by
  means of event-triggered sampling algorithms,'' \emph{{IMA} Journal of
  Mathematical Control and Information}, vol.~31, pp. 415--433, 2014.

\bibitem{Marchand:2013}
N.~Marchand, S.~Durand, and J.~Castellanos, ``A general formula for event-based
  stabilization of nonlinear systems,'' \emph{IEEE Trans. Autom. Control},
  vol.~58, no.~5, pp. 1332--1337, 2013.

\bibitem{Marchand:2013IFAC}
N.~Marchand, J.~Martinez, S.~Durand, and J.~Guerrero-Castellanos, ``Lyapunov
  event-triggered control: a new event strategy based on the control,''
  \emph{IFAC Proceed. Volum.}, vol.~46, no.~23, pp. 324--328, 2013.

\bibitem{Himmelberg:1975}
C.~Himmelberg, ``Measurable relations,'' \emph{Fundamenta Mathematicae},
  vol.~87, no.~1, pp. 53--72, 1975.

\bibitem{Sontag:1991}
Y.~Lin and E.~Sontag, ``A universal formula for stabilization with bounded
  controls,'' \emph{Syst. Control Lett.}, vol.~16, pp. 393--397, 1991.

\bibitem{AntaTabuada:2010}
A.~Anta and P.~Tabuada, ``To sample or not to sample: Self-triggered control
  for nonlinear systems,'' \emph{IEEE Trans. Autom. Control}, vol.~55, no.~9,
  pp. 2030--2042, 2010.

\bibitem{MazoAntaTabuada:2010}
M.~Mazo, A.~Anta, and P.~Tabuada, ``An iss self-triggered implementation of
  linear controllers,'' \emph{Automatica}, vol.~46, no.~8, pp. 1310--1314,
  2010.

\bibitem{HeemelsDonkers:2013}
W.~Heemels and M.~Donkers, ``Model-based periodic event-triggered control for
  linear systems,'' \emph{Automatica}, vol.~49, no.~3, pp. 698--711, 2013.

\bibitem{ProCao:2017Oscill}
A.~Proskurnikov and M.~Cao, ``Synchronization of pulse-coupled oscillators and
  clocks under minimal connectivity assumptions,'' \emph{IEEE Trans. Autom.
  Control}, vol.~62, no.~11, pp. 5873 -- 5879, 2017.

\bibitem{AmesZheng:2006}
A.~Ames, H.~Zheng, R.~Gregg, and S.~Sastry, ``Is there life after {Z}eno?
  {T}aking executions past the breaking ({Z}eno) point,'' in \emph{Proc.
  American Control Conference (ACC)}, 2006, pp. 2652--2657.

\bibitem{HsuSastry1987}
P.~Hsu and S.~Sastry, ``The effect of discretized feedback in a closed loop
  system,'' in \emph{IEEE Conference on Decision and Control}, vol.~26, 1987,
  pp. 1518--1523.

\bibitem{Burlion2006}
L.~Burlion, T.~Ahmed-Ali, and F.~Lamnabhi-Lagarrigue, ``On the stability of a
  class of nonlinear hybrid systems,'' \emph{Nonlinear Analysis: Theory,
  Methods \& Applications}, vol.~65, no.~12, pp. 2236 -- 2247, 2006.

\bibitem{Owens1990}
D.~Owens, Y.~Zheng, and S.~Billings, ``Fast sampling and stability of nonlinear
  sampled-data systems: {P}art~1. existence theorems,'' \emph{IMA J. Math.
  Control Inform.}, vol.~7, pp. 1--11, 1990.

\bibitem{BhatBernstein2000}
S.~Bhat and D.~Bernstein, ``Finite-time stability of continuous autonomous
  systems,'' \emph{SIAM Journal on Control and Optimization}, vol.~38, no.~3,
  pp. 751--766, 2000.

\bibitem{MOULAY2006}
E.~Moulay and W.~Perruquetti, ``Finite time stability and stabilization of a
  class of continuous systems,'' \emph{Journal of Mathematical Analysis and
  Applications}, vol. 323, no.~2, pp. 1430 -- 1443, 2006.

\bibitem{HeemelsDonkersTeel:2013}
W.~Heemels, M.~Donkers, and A.~Teel, ``Periodic event-triggered control for
  linear systems,'' \emph{IEEE Trans. Autom. Control}, vol.~58, no.~4, pp.
  847--861, 2013.

\bibitem{Babu:2016}
F.~Mulakkal-Babu, M.~Wang, B.~van Arem, and R.~Happee, ``Design and analysis of
  full range adaptive cruise control with integrated collision a voidance
  strategy,'' in \emph{Proc. Int. Conf. Intelligent Transp. Syst. (ITSC)},
  2016, pp. 308--315.

\bibitem{WangRiender:2016}
M.~Wang, S.~Hoogendoorn, W.~Daamen, B.~van Arem, B.~Shyrokau, and R.~Happee,
  ``Delay-compensating strategy to enhance string stability of adaptive cruise
  controlled vehicles,'' \emph{Transportmetrica B: Transport Dynamics}, publ.
  online under DOI 10.1080/21680566.2016.1266973.

\bibitem{Chien:1995}
C.-C. Chien, Y.~Zhang, and M.~Lai, ``Regulation layer controller design for
  automated highway systems,'' \emph{Math. Comput. Modeling}, vol.~22, no. 4-7,
  pp. 305--327, 1995.

\bibitem{DolkPloegHeemels:2017}
V.~Dolk, J.~Ploeg, and W.~Heemels, ``Event-triggered control for string-stable
  vehicle platooning,'' \emph{IEEE Trans. Intelligent Transportation Syst.},
  vol.~18, no.~12, pp. 3486--3500, 2017.

\bibitem{AntaTabuada:2008}
A.~Anta and P.~Tabuada, ``Self-triggered stabilization of homogeneous control
  systems,'' in \emph{Proc. American Control Conf.}, 2008, pp. 4129--4134.

\bibitem{McConley:1998}
M.~McConley, M.~Dahleh, and E.~Feron, ``Polytopic control {L}yapunov functions
  for robust stabilization of a class of nonlinear systems,'' \emph{Syst.
  Control Lett.}, vol.~34, pp. 77--83, 1998.

\end{thebibliography}

\appendices

\section{Proofs of Lemmas~\ref{lem.key-lemma} and~\ref{lem.key-lemma1}}\label{app.proof}

Henceforth Assumptions~\ref{ass.F}-\ref{ass.non-degen} are supposed to hold.
For $u_*=\u(x_*)$ and $\xi_0\in B(x_*)$, consider the solution $\xi(t)=\xi(t|\xi_0,u_*)$ to the Cauchy problem~\eqref{eq.cauchy}.
Let $t_*=t_*(\xi_0,u_*)>0$ stand for the first instant $t$ when $W(\xi(t),u_*)=-\sigma\gamma(V(\xi(t))$ and $\Delta_*=\Delta_*(\xi_0,u_*)=[0,t_*]$.
If such an instant does not exist, we put $t_*=\infty$ and $\Delta_*=[0,\infty)$.
Due to Proposition~\ref{prop.tech}, the solution $\xi(t)$ exists on $\Delta_*$ and $\xi(t)\in B(\xi_0)$.

\begin{prop}\label{prop.estimate}
For any $x_*\in\r^d$, $\xi_0\in B(x_*)$ and $u_*=\u(x_*)$, the solution $\xi(t)=\xi(t|\xi_0,u_*)$ satisfies the inequalities:
\be\label{eq.xi-prop}
\begin{gathered}
|\xi(t)-\xi_0|\le c(t,x_*)|F(\xi_0,u_*)|,\\ |F(\xi(t),u_*)|\le (1+\vk(x_*)c(t,x_*))|F(\xi_0,u_*)|,\\
c(t,x_*)\dfb \left(\frac{e^{(2\vk(x_*)+1)t}-1}{2\vk(x_*)+1}\right)^{1/2}.
\end{gathered}
\ee
Here $\vk(x_*)$ is the Lipschitz constant~\eqref{eq.kappa} and $t\in\Delta_*(\xi_0,u_*)$.
\end{prop}
\begin{proof}
Let $\alpha(t)\dfb |\xi(t)-\xi_0|^2/2$. By noticing that $\dot\alpha(t)=(\xi(t)-\xi_0)^{\top}F(\xi(t),u_*)$, one arrives at the inequality
\[
\begin{aligned}
&\dot\alpha(t)=(\xi(t)-\xi_0)^{\top}[F(\xi(t),u_*)-F(\xi_0,u_*)]+\\
&+(\xi(t)-\xi_0)^{\top}F(\xi_0,u_*)\le 2\vk(x_*)\alpha(t)+\alpha(t)+\frac{|F(\xi_0,u_*)|^2}{2}
\end{aligned}
\]
(by assumption, $\xi_0\in B(x_*)$ and thus $\xi(t)\in B(x_*)\,\forall t\in\Delta_*(\xi_0,u_*)$).
The usual comparison lemma~\cite{Khalil} implies that $\alpha(t)\le\beta(t)$, where $\beta(t)$ is the solution to the Cauchy problem
\[
\dot\beta(t)=[2\vk(x_*)+1]\beta(t)+\frac{|F(\xi_0,u_*)|^2}{2},\quad \beta(0)=\alpha(0)=0.
\]
A straightforward computation shows that $\beta(t)=c(t,x_*)^2|F(\xi_0,u_*)|^2/2$, which entails the the first inequality in~\eqref{eq.xi-prop}.
The second inequality is immediate from~\eqref{eq.kappa} since $|F(\xi(t),u_*)|\le |F(\xi_0,u_*)|+\vk(x_*)|\xi(t)-\xi_0|$.
\end{proof}

To simplify the estimates for the minimal dwell time, we will use the following simple inequality for the function $c(t,x_*)$.
\begin{prop}
If $0\le t\le (1+2\vk(x_*))^{-1}<1$, then
\be\label{eq.c-est}
c(t,x_*)\le \sqrt{te}\le \sqrt{e}.
\ee
\end{prop}
\begin{proof}\renewcommand{\IEEEQED}{}
Denoting for brevity $\vk=\vk(x_*)$, the statement follows from the mean value theorem, applied to $e^{(2\vk+1)t}$:
\ben
\begin{gathered}
\exists t_0\in (0,t):\; e^{(2\vk+1)t}-1=t(2\vk+1)e^{(2\vk+1)t_0}\leq (2\vk+1)e,\\
c(t,x_*)^2=\frac{e^{(2\vk+1)t}-1}{2\vk+1}=te^{(2\vk+1)t_0}\leq e.\quad\quad\quad\quad\quad\quad\IEEEQEDclosed
\end{gathered}
\een
\end{proof}
\begin{corollary}
Let $\xi_0\in B(x_*)$, $u_*=\u(x_*)$ and $\xi(t)=\xi(t|\xi_0,u_*)$, where $t\in\Delta_*(\xi_0,u_*)\cap \left[0,(1+2\vk(x_*))^{-1}\right]$. Then
\begin{gather}
\begin{aligned}
|W(&\xi(t),u_*)-W(\xi_0,u_*)|\le \\&\le\sqrt{t}\mu(x_*)\left(|V'(\xi_0)|\,|F(\xi_0,u_*)|+|F(\xi_0,u_*)|^2\right),\\
\end{aligned}\label{eq.delta_w1}\\
\mu(x_*)\dfb \sqrt{e}\max\left\{\vk(x_*),\nu(x_*)(1+\vk(x_*)\sqrt{e})\right\}.\label{eq.def-mu}
\end{gather}
Here $\nu(x_*)$ is the Lipschitz constant from~\eqref{eq.nu}.
\end{corollary}
\begin{proof}\renewcommand{\IEEEQED}{}
Recalling that $\xi=\xi(t)\in B(\xi_*)$, one has
\[
\begin{aligned}
|W(\xi,&u_*)-W(\xi_0,u_*)|\leq |\left(V'(\xi)-V'(\xi_0)\right)F(\xi,u_*)|+\\
+&|V'(\xi_0)\left(F(\xi,u_*)-F(\xi_0,u_*)\right)|\overset{\eqref{eq.kappa},\eqref{eq.nu}}{\leq}\\
\leq & \nu(x_*)|\xi-\xi_0| |F(\xi,u_*)|+\vk(x_*)|V'(\xi_0)| |\xi-\xi_0|\overset{\eqref{eq.xi-prop}}{\leq}\\
\leq & \nu(x_*)c(t,x_*)(1+\vk(x_*)c(t,x_*))|F(\xi_0,u_*)|^2+\\
+ & \vk(x_*)c(t,x_*)|V'(\xi_0)|\,|F(\xi_0,u_*)|\overset{\eqref{eq.c-est}}{\leq}\\
\leq & \sqrt{te}\nu(x_*)(1+\vk(x_*)\sqrt{e})|F(\xi_0,u_*)|^2+\\
+&\sqrt{te}\vk(x_*)|V'(\xi_0)|\,|F(\xi_0,u_*)|\overset{\eqref{eq.def-mu}}{\leq}\\
\leq &\sqrt{t}\mu(x_*)\left(|V'(\xi_0)|\,|F(\xi_0,u_*)|+|F(\xi_0,u_*)|^2\right).\quad\quad\quad\IEEEQEDclosed
\end{aligned}
\]
\end{proof}

\subsection{The proof of Lemma~\ref{lem.key-lemma}}

In this subsection, $\xi(t)=\xi(t|x_*,\u(x_*))$ stands for the solution of the special Cauchy problem~\eqref{eq.cauchy} with $\xi_0=x_*$.
For brevity, let $t_*(x_*)\dfb t_*(x_*,u_*)$ and $\Delta(x_*)\dfb \Delta(x_*,u_*)$.
To construct $\tau(\cdot)$, introduce an auxiliary function
\be\label{eq.tau}
\tilde\tau_{\sigma}(x_*)\dfb\min\left\{\frac{(1-\sigma)^2}{\mu(x_*)^2M(x_*)^2},\frac{1}{1+2\vk(x_*)}\right\}>0.
\ee
Besides this, in the case where $\gamma\in C^1$ (and the monotonicity of $\gamma$ is not supposed) we consider an additional function
\be\label{eq.tau1}
\begin{gathered}
\hat\tau_{\sigma}(x_*)\dfb\min\left(\tilde\tau_{\sigma_0}(x_*),\frac{\sigma_0-\sigma}{\sigma(2-\sigma_0)\rho(x_*)}\right),\\
\rho(x_*)\dfb\max_{0\le v\le V(x_*)}\min\{0,-\gamma'(v)\},\quad \sigma_0\dfb \frac{1+\sigma}{2}.
\end{gathered}
\ee
We now introduce $\tau(x)$ as follows
\be\label{eq.tau-final}
\tau(x_*)\dfb
\begin{cases}
\tilde\tau_{\sigma}(x_*),\quad &\text{$\gamma$ is non-decreasing}\\
\hat\tau_{\sigma}(x_*),\quad &\text{otherwise}.
\end{cases}
\ee
It can be easily shown that $\tau(\cdot)$ is uniformly positive on any compact set. If the functions $\vk,\nu,M,\rho$ are \emph{globally} bounded,
the same holds for $\mu$, and thus $\tau(\cdot)$ is \emph{uniformly} positive.

To prove Lemma~\ref{lem.key-lemma}, it suffices to show that $t_*(x_*)\ge\tau(x_*)$.
For $x_*=0$, $t_*(x_*)=\infty$ and the statement is obvious, otherwise for any $t\in\Delta_*(x_*)\cap [0,(1+2\vk(x_*))^{-1})$ one has
\[
\begin{aligned}
|W(\xi(t),u_*)-W(x_*,u_*)|\overset{\eqref{eq.non-degen1},\eqref{eq.delta_w1}}{\le}\sqrt{t}\mu(x_*)M(x_*)|W(x_*,u_*)|
\end{aligned}
\]
(recall that $W(x_*,u_*)=V'(x_*)\bar F(x_*)$). For $t<\tilde\tau_{\sigma}(x_*)$, one has $\sqrt{t}\mu(x_*)M(x_*)<1-\sigma$. Hence, on the interval
$t\in\Delta_*(x_*)\cap [0,\tilde\tau_{\sigma}(x_*))$ the following inequalities hold
\begin{gather}
\begin{aligned}
W&(\xi(t),u_*)<W(x_*,u_*)+(1-\sigma)|W(x_*,u_*)|\\
&=|W(x_*,u_*)|(-1+1-\sigma)=-\sigma|W(x_*,u_*)|=\\
&=\sigma W(x_*,u_*)\overset{\eqref{eq.inf-u-gamma}}{\leq} -\sigma\gamma(V(x_*))
\end{aligned}\label{eq.ineq1}\\
\begin{aligned}
&W(\xi(t),u_*)>W(x_*,u_*)-(1-\sigma)|W(x_*,u_*)|\\
&=|W(x_*,u_*)|(-1-1+\sigma)=(\sigma-2)|W(x_*,u_*)|=\\
&=(2-\sigma) W(x_*,u_*).
\end{aligned}\label{eq.ineq2}
\end{gather}

Consider first the case where $\gamma$ is non-decreasing. Since $V(\xi(t))\overset{\eqref{eq.ineq1}}{\leq} V(x_*)$ and $\gamma(V(x_*))\ge \gamma(V(\xi(t)))$, one has
\[
W(\xi(t),u_*)\overset{\eqref{eq.ineq1}}{<} -\sigma\gamma(V(\xi(t)))\quad \forall t\in\Delta_*(x_*)\cap [0,\tilde\tau_{\sigma}(x_*)).
\]
By definition of $t_*$, we have $\Delta_*(x_*)\cap [0,\tilde\tau_{\sigma}(x_*))\subseteq [0,t_*(x_*))$, that is, $t_*(x_*)\ge\tilde\tau_{\sigma}(x_*)=\tau(x_*)$, which finishes the proof.

In the case of $\gamma\in C^1$, choose any $t\in\Delta(x_*)\cap [0,\tau(x_*))$. Due to the mean-value theorem, $\delta_t\in (0,t)$ exists such that
\[
\begin{aligned}
&\gamma(V(\xi(t)))-\gamma(V(x_*))=t\gamma'(V(\xi(\delta_t)))W(\xi(\delta_t),u_*)=\\
&=t|W(\xi(\delta_t),u_*)|(-\gamma'(V(\xi(\delta_t))))\overset{\eqref{eq.tau1}}{\le} t\rho(x_*)|W(\xi(\delta_t),u_*)|.
\end{aligned}
\]
The latter inequality holds due to the definition of $\rho(\cdot)$ in~\eqref{eq.tau1} since $V(\xi(\delta_t))\le V(x_*)$.
Applying now~\eqref{eq.ineq2} to $\sigma=\sigma_0$ and recalling that $t<\tilde\tau_{\sigma_0}(x_*)$, one shows that $|W(\xi(\delta_t),u_*)|\le (2-\sigma_0)|W(x_*,u_*)|$. Since $\gamma(V(x_*))\overset{\eqref{eq.inf-u-gamma}}{\leq}|W(x_*,u_*)|$,
\[
\begin{gathered}
\gamma(V(\xi(t)))\leq |W(x_*,u_*)|(1+t\rho(x_*)(2-\sigma_0))\leq \\
\leq |W(x_*,u_*)|\left(1+\hat\tau_{\sigma}(x_*)\rho(x_*)(2-\sigma_0)\right)\overset{\eqref{eq.tau1}}{\leq} \\ \leq \sigma^{-1}\sigma_0|W(x_*,u_*)|.
\end{gathered}
\]
Using the inequality~\eqref{eq.ineq1} with $\sigma_0$ instead of $\sigma$, one arrives at
\[
\begin{gathered}
W(\xi(t),u_*)<\sigma_0W(x_*,u_*)\leq -\sigma\gamma(V(\xi(t))).
\end{gathered}
\]
Therefore, $\Delta_*(x_*)\cap [0,\tau(x_*))\subseteq [0,t_*(x_*))$, and hence $t_*(x_*)\ge\tau(x_*)$, which finishes the proof of Lemma~\ref{lem.key-lemma}.$\blacksquare$

\subsection{Proof of Lemma~\ref{lem.key-lemma1}}

In this subsection, we deal with a more general Cauchy problem~\eqref{eq.cauchy}, where $u_*=\u(x_*)$, but $\xi_0=\bar x\ne x_*$; it is only assumed that
that $\bar x\in B(x_*)$. The proof follows the same line as the proof of Lemma~\ref{lem.key-lemma} and employs the function
\begin{gather}\label{eq.tau2}
\bar\tau_{\sigma,\tilde\sigma,K}(x_*)\dfb\min\left\{\frac{(\tilde\sigma-\sigma)^2}{K^2\mu(x_*)^2M(x_*)^2\tilde\sigma^2}
,\frac{1}{1+2\vk(x_*)}\right\}
\end{gather}
and, in the case where $\gamma\in C^1$, the function
\be\label{eq.tau2+}
\begin{gathered}
\breve\tau_{\sigma,\tilde\sigma,K}(x_*)\dfb\min\left(\bar\tau_{\sigma_1,\tilde\sigma,K}(x_*),\frac{\sigma_1-\sigma}{\sigma(2\tilde\sigma-\sigma_1)\rho(x_*)}\right)\\
\sigma_1\dfb\frac{\tilde\sigma+\sigma}{2}.
\end{gathered}
\ee
Similar to~\eqref{eq.tau-final}, we define the function $\tau_{\sigma,\tilde\sigma,K}$ as follows
\[
\tau^0(x_*)\dfb
\begin{cases}
\bar\tau_{\sigma,\tilde\sigma,K}(x_*),\quad &\text{$\gamma$ is non-decreasing}\\
\breve\tau_{\sigma,\tilde\sigma,K}(x_*),\quad &\text{$\gamma\in C^1$}.
\end{cases}
\]

We are going to show that $t_*(\bar x,u_*)\ge\tau^0(x_*)$ when $\bar x\in B(x_*)$ and $P(\bar x,u_*)$ is true.
Using the inequality
\be\label{eq.tech1}
|V'(\bar x)|\,|F(\bar x,u_*)|+|F(\bar x,u_*)|^2\overset{P(\bar x,u_*)}{\leq} K|W(\bar x,u_*)|,
\ee
one shows that for any $t\in\Delta_*(\bar x,u_*)\cap [0,(1+2\vk(x_*))^{-1})$
\ben
\begin{aligned}
|W(\xi(t),u_*)-W(\bar x,u_*)|\overset{\eqref{eq.delta_w1},\eqref{eq.tech1}}{\le}\sqrt{t}K\mu(x_*)M(x_*)|W(\bar x,u_*)|.
\end{aligned}
\een
For any $t\in\Delta_*(\bar x,u_*)\cap [0,\bar\tau_{\sigma,\tilde\sigma,K}(x_*))$ one has $\sqrt{t}K\mu(x_*)M(x_*)<1-\sigma\tilde\sigma^{-1}$,  which allows to prove the following counterparts
of the inequalities~\eqref{eq.ineq1} and~\eqref{eq.ineq2}
\begin{gather}
\begin{aligned}
W(\xi(t),u_*)<W(\bar x,u_*)+(1-\sigma\tilde\sigma^{-1})|W(\bar x,u_*)|=\\=\sigma\tilde\sigma^{-1} W(\bar x,u_*)\overset{P(\bar x,u_*)}{\leq} -\sigma\gamma(V(\bar x)),
\end{aligned}\label{eq.ineq1+}\\
\begin{aligned}
W(\xi(t),u_*)>W(\bar x,u_*)-(1-\sigma\tilde\sigma^{-1})|W(\bar x,u_*)|=\\=(2-\sigma\tilde\sigma^{-1})W(\bar x,u_*).
\end{aligned}\label{eq.ineq2+}
\end{gather}

In the first case, where $\gamma$ is non-decreasing, the inequality~\eqref{eq.ineq1+} implies that $W(\xi(t),u_*)<-\sigma\gamma(V(\xi(t)))$ whenever $t\in\Delta_*(\bar x,u_*)\cap [0,\bar\tau_{\sigma,\tilde\sigma,K}(x_*))$ since $V(\xi(t))\le V(\bar x)$. This implies that $t_*(\bar x,u_*)\ge \bar\tau_{\sigma,\tilde\sigma,K}(x_*)=\tau^0(x_*)$.

In the case of $\gamma\in C^1$, the mean value theorem implies that
\[
\begin{aligned}
\gamma&(V(\xi(t)))-\gamma(V(\bar x))=t\gamma'(V(\xi(\delta_t)))W(\xi(\delta_t),u_*)=\\
&=t|W(\xi(\delta_t),u_*)|(-\gamma'(V(\xi(\delta_t))))\leq t\rho(x_*)|W(\xi(\delta_t),u_*)|.
\end{aligned}
\]
The latter inequality holds due to the definition of $\rho(x_*)$ in~\eqref{eq.tau1} since $V(\xi(\delta_t))\le V(\bar x)\le V(x_*)$.
Applying~\eqref{eq.ineq2+} to $\sigma=\sigma_1$, one shows that $|W(\xi(\delta_t),u_*)|\le (2-\sigma_1\tilde\sigma^{-1})|W(\bar x,u_*)|$
whenever $t\le\breve\tau_{\sigma_1,\tilde\sigma,K}(x_*)$. The condition $P(\bar x,u_*)$ implies that $\gamma(V(\bar x))\le\tilde\sigma^{-1}|W(\bar x,u_*)|$. Hence
for any $t\in\Delta_*(\bar x,u_*)\cap [0,\tau^0(x_*))$ one obtains that
\[
\begin{aligned}
\gamma(V(\xi(t)))\leq |W(\bar x,u_*)|(\tilde\sigma^{-1}+t\rho(x_*)(2-\sigma_1\tilde\sigma^{-1}))<\\
< |W(\bar x,u_*)|\left(\tilde\sigma^{-1}+\breve\tau_{\sigma_1,\tilde\sigma,K}(x_*)\rho(x_*)(2-\sigma_1\tilde\sigma^{-1})\right)\overset{\eqref{eq.tau2+}}{\leq}\\
\leq \tilde\sigma^{-1}\sigma^{-1}\sigma_1|W(\bar x,u_*)|.
\end{aligned}
\]
Using the inequality~\eqref{eq.ineq1+} for $\sigma_1$, one arrives at
\[
\begin{gathered}
W(\xi(t),u_*)<\sigma_1\tilde\sigma^{-1}W(\bar x,u_*)\leq -\sigma\gamma(V(\xi(t))).
\end{gathered}
\]
This implies that $t_*(\bar x,u_*)\ge \breve\tau_{\sigma,\tilde\sigma,K}(x_*)=\tau^0(x_*)$, which finishes the proof of Lemma~\ref{lem.key-lemma1} in the second case.


\section{Discussion on Assumption~\ref{ass.non-degen}}\label{app.discuss}

Assumption~\ref{ass.non-degen} complements the Lyapunov inequality~\eqref{eq.inf-u-gamma} in the following sense.
Decompose the right-hand side of the continuous-time system $\bar F(x)=F(x,\u(x))$ into the sum of two vectors, one parallel to the CLF's gradient $\nabla V(x)=V'(x)^{\top}$ and the other orthogonal to it
\[
\bar F(x)=-\alpha(x)\nabla V(x)+v_{\perp}(x),
\]
where $\alpha(x)\in\r$ and $\nabla V(x)\perp v_{\perp}(x)\in\r^d\,\forall x\ne 0$.
The Lyapunov inequality~\eqref{eq.inf-u-gamma} gives a lower bound for $\alpha(x)$:
\be\label{eq.alpha-tech}
\alpha(x)\ge \frac{\gamma(V(x))}{|V'(x)|^2},
\ee
but neither specifies any \emph{upper} bound on $\alpha$, nor restricts the transverse component $v_{\perp}(x)$ in any way.
The definition does not exclude fast-oscillating solutions, changing much faster than the CLF is decaying
$|\dot x(t)|=|\bar F(x(t))|\gg |\dot V(x(t))|$. This happens e.g. when the orthogonal component $v_{\perp}$ (which influence
$\dot x$, but does not affect $\dot V(x)$) dominates over the parallel component $(\alpha\nabla V)$ or when $\alpha(x)$ grows unbounded when $|x|\to 0$. If the continuous-time control $u(t)=\u(x(t))$ is also fast-changing, it is intuitively clear that
no finite sampling rate can appear sufficient to maintain the prescribed convergence rate (an explicit example is given below). The restrictions of Assumption~\ref{ass.non-degen} prohibit these pathological behaviors and require, first, that the transverse component of the velocity $v_{\perp}$ is proportional to the gradient component $(-\alpha\nabla V)$, and, second, both components decay as $O(|V'(x)|)$ as $|x|\to 0$. Mathematically, this can be formulated as follows.
\begin{prop}\label{lem.tech1}
Assumption~\ref{ass.non-degen} holds if and only if $\alpha(x)$ is locally bounded and $\bar F(x)\le \tilde M(x)\alpha(x)|V'(x)|$,
where $\tilde M$ is a locally bounded function.
\end{prop}
\begin{IEEEproof}
Notice that $|V'(x)\bar F(x)|=\alpha(x)|V'(x)|^2$ and
\[
\max(\alpha(x)|V'(x)|,|v_{\perp}(x)|)\le|\bar F(x)|\le \alpha(x)|V'(x)|+|v_{\perp}(x)|.
\]
The statement now follows from Lemma~\ref{lem.tech}.
\end{IEEEproof}

We now proceed with an example, demonstrating that Assumption~\ref{ass.non-degen} cannot be fully discarded even in the situation of
exponential convergence. Consider a linear planar system
\be\label{eq.system-polar}
\dot x_1=x_2+u_1,\quad \dot x_2=-x_1+u_2,\quad u_1,u_2\in\r.
\ee
Consider now the exponentially stabilizing controller
\[
\begin{pmatrix}
u_1\\u_2
\end{pmatrix}=\u(x)=
-\begin{pmatrix}
x_1\\x_2
\end{pmatrix}+\frac{1}{\sqrt{x_1^2+x_2^2}}\begin{pmatrix}
x_2\\-x_1
\end{pmatrix},\;\;x\ne 0,
\]
and $\u(0)=0$. Obviously, for $V(x)=\frac{1}{2}|x|^2$ and $\gamma(v)=2v$
one has $V'(x)F(x,\bar U(x))=x^{\top}u=-|x|^2=-\gamma(V(x))$, so the continuous-time control exponentially stabilizes the system. Assumption~\ref{ass.non-degen} is violated since
\[
|\bar F(x)|^2=2|x|^2+2|x|+1\xrightarrow[|x|\to 0]{} 1.
\]
We are going to show that algorithm~\eqref{eq.alg1} cannot provide locally uniformly positive dwell-time. To prove this, we introduce the polar coordinates $x_1=r\cos\vp, x_2=r\sin\vp$, rewriting the dynamics~\eqref{eq.system-polar} in the area $\r^2\setminus\{0\}$ as
\be\label{eq.system-polar1}
\begin{gathered}
\begin{aligned}
&\dot r\cos\vp-r\dot\vp\sin\vp=r\sin\vp+u_1\\
&\dot r\sin\vp+r\dot\vp\cos\vp=-r\cos\vp+u_2
\end{aligned}\\
\Updownarrow\\
\begin{aligned}
&\dot r=u_1\cos\vp+u_2\sin\vp\\
&\dot\vp=-1+r^{-1}(u_2\cos\vp-u_1\sin\vp)
\end{aligned}
\end{gathered}
\ee
Suppose that the algorithm starts at some point $x_*=r_*(\cos\vp_*,\sin\vp_*)^{\top}$ with $r_*=|x_*|\in (0,1)$, and the initial control input is $u_*=\u(x_*)$. On the interval $(0,t_1)$, where $t_1$ stands for the instant of first event, one has
\be\label{eq.system-polar2}
\begin{aligned}
&\dot r=-r_*\cos(\vp_*-\vp)+\sin(\vp_*-\vp)\\
&\dot\vp=-1+r^{-1}r_*\sin(\vp_*-\vp)-r^{-1}\cos(\vp_*-\vp)
\end{aligned}
\ee
By definition of $t_1$, the CLF $V(x)=|x|^2=r^2$ decays on $[0,t_1)$, and thus $\dot r(t)\le 0$ and $r(t)\le r_*$. When $\vp(t)$ is close to $\vp_*$, one obviously has $\dot\vp\le -1$ since $r_*\sin(\vp_*-\vp)<\cos(\vp_*-\vp)$. Therefore, $\vp(t)<\vp_*$ for any $t\in(0,t_1]$. Since $\dot r\le 0$, one has $\sin(\vp_*-\vp)\le r_*\cos(\vp_*-\vp)$, thus
\begin{gather}
0<\vp_*-\vp(t)\le\arctan r_*\label{eq.aux-1}\\
\begin{gathered}
\sin(\vp_*-\vp(t))\le\frac{r_*}{\sqrt{1+r_*^2}}\le r_*\cos(\vp_*-\vp(t)),
\end{gathered}\label{eq.aux-2}
\end{gather}
on $(0,t_1]$ (inequalities~\eqref{eq.aux-2} are based on~\eqref{eq.aux-1} and the decreasing/increasing of $\cos/\sin$ respectively on $[0,\pi)$). Hence
\[
\dot\vp\overset{\eqref{eq.system-polar1}}{\le}
-1+\frac{r_*^{2}-1}{r\sqrt{1+r_*^2}}\overset{r\le r_*<1}{\le} -1-\frac{1-r_*^2}{r_*\sqrt{1+r_*^2}}
\]
on $(0,t_1]$, which entails, accordingly to~\eqref{eq.aux-1}, that
\[
t_1\le \frac{r_*\sqrt{1+r_*^2}\arctan r_*}{r_*\sqrt{1+r_*^2}+1-r_*^2}\xrightarrow[r_*\to 0]{}0.
\]
Therefore, the algorithm does not provide local uniform positivity of the dwell-time (this algorithm in fact exhibits Zeno behavior, but the proof is omitted due to the page limit).
\color{black} 

\begin{IEEEbiography}
{Anton Proskurnikov}(M'13, SM'18)
was born in St. Petersburg, Russia, in 1982. He received the
M.Sc. (``Specialist'') and Ph.D. (``Candidate of Sciences'') degrees in applied mathematics from St. Petersburg State University in 2003 and 2005,
respectively.

Anton Proskurnikov is currently a Researcher at Delft Center for Systems and Control, Delft University of Technology (TU Delft), The Netherlands. Before joining TU Delft,
he stayed with St. Petersburg State University (2003-2010) as an Assistant Professor and the University of Groningen (2014-2016) as a postdoctoral researcher.
He also occupies part-time research positions at Institute for Problems of Mechanical Engineering of the Russian Academy of Sciences and ITMO University.
His research interests include dynamics of complex networks, robust and nonlinear control, optimal control and control applications to social
and biological sciences. He is a member of Editorial Board of the Journal of Mathematical Sociology.
\end{IEEEbiography}
\begin{IEEEbiography}
{Manuel Mazo Jr.} (S'99, M'11, SM'18) is an associate professor at the Delft Center for Systems and Control, Delft University of Technology (The Netherlands). He received the M.Sc. and Ph.D.degrees in Electrical Engineering from the University of California, Los Angeles, in 2007 and 2010 respectively. He also holds a Telecommunications Engineering "Ingeniero" degree from the Polytechnic University of Madrid (Spain), and a "Civilingenj\"or" degree in Electrical Engineering from the Royal Institute of Technology (Sweden), both awarded in 2003. Between 2010 and 2012 he held a joint post-doctoral position at the University of Groningen and the (now defunct) innovation centre INCAS3, The Netherlands. His main research interest is the formal study of problems emerging in modern control system implementations, in particular, the study of networked control systems and the application of formal verification and synthesis techniques to control. He has been the recipient of a University of Newcastle Research Fellowship (2005), the Spanish Ministry of Education/UCLA Fellowship (2005-2009), the Henry Samueli Scholarship from the UCLA School of Engineering and Applied Sciences (2007/2008) and ERC Starting Grant (2017).
\end{IEEEbiography}

\end{document}